\documentclass[final,comsoc]{IEEEtran}
\usepackage[T1]{fontenc}
\usepackage[utf8]{inputenc}
\usepackage{amsmath}
\usepackage{cases} 
\usepackage[cmintegrals]{newtxmath} 
\usepackage{bm} 
\usepackage[caption=false,font=footnotesize]{subfig}
\usepackage{booktabs} 
\usepackage{xcolor}
\let\ieeeappendix\appendix
\usepackage[capitalize]{cleveref} 
\crefformat{equation}{(#2#1#3)}
\crefrangeformat{equation}{(#3#1#4) to~(#5#2#6)}
\crefmultiformat{equation}{(#2#1#3)}%
{ and~(#2#1#3)}{, (#2#1#3)}{ and~(#2#1#3)}
\let\appendix\ieeeappendix
\usepackage[nolist]{acronym} 
\usepackage{siunitx} 
\usepackage{cite}
\usepackage{microtype} 
\usepackage{graphicx}
\usepackage{marginnote}

\graphicspath{{graphics/}}

\DeclareMathOperator{\erf}{erf} 
\DeclareMathOperator{\e}{e} 
\DeclareMathOperator{\Langevin}{L} 
\DeclareMathOperator{\Dz}{\frac{\partial}{\partial z}}
\DeclareMathOperator{\E}{E}
\DeclareSIUnit{\bit}{bit}

\newtheorem{theorem}{Theorem}

\newcommand{\mathsc}[1]{{\normalfont\textsc{#1}}}

\providecommand\diff{\,\mathrm{d}}
\newcommand\D[2][]{\frac{\partial^{#1}}{\partial #2^{#1}}}
\newcommand{\crit}{\mathrm{crit}} 
\newcommand{\im}{\mathrm{j}} 
\newcommand{\hyd}{\mathrm{h}} 
\newcommand{\f}{\mathrm{f}} 
\newcommand{\m}{\mathrm{m}} 
\newcommand{\TX}{\mathsc{tx}} 
\newcommand{\RX}{\mathsc{rx}} 
\newcommand{\ob}{\mathrm{ob}} 
\newcommand{\sat}{\mathrm{s}} 
\newcommand{\kb}{k_\mathsc{b}} 
\newcommand{\Tf}{T_\f} 
\newcommand{\vm}{v_\m} 
\newcommand{\vf}{v_\f} 
\newcommand{\ucrit}{u_\crit} 
\newcommand{\Rmag}{R_\mathrm{mag}} 
\newcommand{\Lmag}{L_\mathrm{mag}} 
\newcommand{\ka}{k_\mathrm{a}} 
\providecommand{\adcoeff}{k_\textrm{a}} 
\providecommand{\mcoeff}{\kappa} 
\providecommand{\madcoeff}{k_\Delta} 
\providecommand{\Pad}{P_\textrm{ad}} 
\providecommand{\norm[1]}{\lVert #1 \rVert} 
\providecommand{\Msat}{M_\sat} 
\providecommand{\meanx}{\overline{x}} 
\providecommand{\Pobx}{P_{\textrm{ob},x}} 
\providecommand{\Poby}{P_{\textrm{ob},y}} 
\providecommand{\Pobz}{P_{\textrm{ob},z}} 
\providecommand{\Cs}{C_\textrm{s}}
\providecommand{\Ns}{N_\textrm{s}}
\providecommand{\Rhyd}{R_\textrm{h}}
\providecommand{\Pobyeq}{P^\infty_{\ob,y}}
\providecommand{\Pobzeq}{P^\infty_{\ob,z}}

\providecommand{\Lmag}{L_\textrm{mag}}
\providecommand{\Rmag}{R_\textrm{mag}}
\providecommand{\dm}{d_\textrm{m}}
\providecommand{\Ntx}{N_\textsc{tx}}

\providecommand{\mh}{\mu_\textrm{h}}
\providecommand{\sh}{\sigma_\textrm{h}}
\providecommand{\Vs}{V_\textrm{s}}
\providecommand{\Vhyd}{V_\textrm{h}}
\providecommand{\Fm}{F_\textrm{m}}
\providecommand{\Pobeq}{P_\textrm{ob}^\infty}

\providecommand{\Pob}{P_\textrm{ob}}
\providecommand{\Nobmean}{\overline{N}_\textrm{ob}}

\providecommand{\PoissonCDF}{F}
\providecommand{\Zint}{\widetilde{Z}}
\providecommand{\Yint}{\widetilde{Y}}
\providecommand{\vmnom}{v_{\textrm{m},0}}
\providecommand{\de}{\,\mathrm{d}}
\providecommand{\nRXmean}{\overline{n}_\textsc{rx}}
\providecommand{\sy}{o}

\newcommand{\changed}[2][]{{#2}}

\begin{document}
\title{
    Magnetic Nanoparticle Based Molecular Communication in Microfluidic Environments
}

\author{%
    Wayan~Wicke,~\IEEEmembership{Student Member,~IEEE,}
    Arman~Ahmadzadeh,~\IEEEmembership{Student Member,~IEEE,}
    Vahid~Jamali,~\IEEEmembership{Student Member,~IEEE,}
    Harald~Unterweger,
    Christoph~Alexiou,
    and Robert~Schober,~\IEEEmembership{Fellow,~IEEE}%
    \vspace*{-5mm}%
    \thanks{%
        This work was supported by the Emerging Fields Initiative (EFI) of the Friedrich-Alexander-Universit\"at Erlangen-N\"urnberg (FAU) and the STAEDTLER Stiftung.
        This work was presented in part at IEEE WCNC 2018 \cite{wicke_molecular_2017}.
        \textit{(Corresponding author: Wayan Wicke.)}
    }%
    \thanks{%
        W.~Wicke, A.~Ahmadzadeh, V.~Jamali, and R.~Schober are with the Institute for Digital Communications at the Friedrich-Alexander-Universit\"at Erlangen-N\"urnberg (FAU), 91058 Erlangen, Germany (e-mail: wayan.wicke@fau.de; arman.ahmadzadeh@fau.de; vahid.jamali@fau.de; robert.schober@fau.de).
    }%
    \thanks{%
        H.~Unterweger and C.~Alexiou are with the Section for Experimental Oncology and Nanomedicine (SEON), Universit\"atsklinikum Erlangen, 91012 Erlangen, Germany  (email: harald.unterweger@uk-erlangen.de; christoph.alexiou@uk-erlangen.de).%
    }%
}%

\maketitle%
\begin{abstract}
The possibility to guide and control magnetic nanoparticles in a non-invasive manner has spawned various applications in biotechnology such as targeted drug delivery and sensing of biological substances.
These applications are facilitated by the engineering of the size, selective chemical reactivity, and general chemical composition of the employed particles.
Motivated by their widespread use and favorable properties, in this paper, we 
\changed[II:6]{provide a theoretical study of the potential benefits of magnetic nanoparticles for the design of molecular communication systems.}
In particular, we consider magnetic nanoparticle based communication in a microfluidic channel where an external magnetic field is employed to attract the information-carrying particles to the receiver.
We show that the particle transport affected by Brownian motion, fluid flow, and an external magnetic field can be mathematically modeled as diffusion with drift.
Thereby, we reveal that the key parameters determining the magnetic force are the particle size and the magnetic field gradient.
Moreover, we derive an analytical expression for the channel impulse response, which is used to evaluate the potential gain in the expected number of observed nanoparticles due to the magnetic field.
Furthermore, adopting the symbol error rate as performance metric, we show that using magnetic nanoparticles
\changed[II:6]{can enable} 
reliable communication in the presence of disruptive fluid flow.
Numerical results obtained by particle-based simulation validate the accuracy of the derived analytical expressions.
\end{abstract}


\begin{acronym}
    \acro{ISI}{intersymbol interference}
    \acro{MC}{Molecular communication}
    \acro{MNP}{magnetic nanoparticle}
    \acro{TX}{transmitter}
    \acro{RX}{receiver}
    \acro{OOK}{on-off keying}
    \acro{ODE}{ordinary differential equation}
    \acro{PDE}{partial differential equation}
    \acro{PDF}{probability density function}
    \acro{SER}{symbol error rate}
\end{acronym}
\section{Introduction}
\ac{MC} is one of the mechanisms that biological cells use to communicate with each other~\cite[Ch.~16]{alberts_essential_2013}.
In natural \ac{MC} systems, information is conveyed by specific patterns of molecule releases, e.g., different numbers or different types of molecules.
Thereby, in
\changed[II:7]{diffusive} 
\ac{MC} environments, the \emph{information molecules} propagate by \emph{Brownian motion} where the movement of particles is caused by thermally induced collisions with the molecules of the embedding liquid.

\changed{On the other hand, recent advances in the field of nanotechnology have enabled the development of small-scale devices, so-called nanomachines. Nanomachines have functional components that are on the order of nanometers in size~(\SI{e-9}{\meter}) and are capable of performing simple computation, sensing, or actuation tasks~\cite{farsad_comprehensive_2016}.}
\changed[II:7]{To perform more complex tasks, multiple nanomachines have to cooperate. As an example, programmed drug bearing cells}
could cooperate and fight a local infection by adjusting the release of a pharmaceutical in a coordinated and controlled manner~\cite[Ch.~8]{nakano_molecular_2013}.
For this smart collaboration of nanomachines, communication is essential.
Thereby, a message might trigger a certain chemical process which, in turn, may cause a desired action at a receiving nanomachine.
For example, a ligand binding to a receptor embedded in the membrane of a receiving cell could initiate a signaling cascade inside the cell which alters its state.
In this context, \ac{MC} has recently attracted considerable attention as a biocompatible approach for synthetic communication at the cellular level.
Communication theoretic frameworks have already proven useful for analyzing natural \ac{MC} systems~\cite{Tostevin_Mutual_2009}.
Moreover, for the design of artificial \ac{MC} networks, basic communication theoretic tools have been developed to achieve robust and efficient signaling~\cite{nakano_molecular_2013}.
\changed[II:4]{However, it is noteworthy that \ac{MC} research is still in its infancy and focused mostly on theoretical aspects with only a few experimental systems available at macroscale \cite{farsad_comprehensive_2016,Unterweger_Experimental_2018,shakya_correlated_2018,mcguiness_experimental_2018}.} 

For \ac{MC}, usually naturally occurring molecules such as proteins are considered as information carriers~\cite[Ch.~2]{nakano_molecular_2013}.
However, apart from the challenges that need to be overcome for realizing synthetic biological \ac{MC} systems at nanoscale~\cite{farsad_comprehensive_2016}, there are also severe inherent limitations. 
In particular, the movement of the information carriers induced by Brownian motion is random and cannot be stirred towards the receiver, i.e., many of the released molecules may not arrive at the receiver.
Moreover, molecules suspended in a fluid are very sensitive to fluid flow which easily dominates the diffusive movement and in many cases (e.g., in blood vessels) cannot be externally controlled.
\changed[I:8]{In fact, fluid flow may further reduce the number of particles observed at the receiver.} 
In this paper, \changed{for the example scenario of a microfluidic channel,} we will show that these problems can be \changed{addressed} by using \acp{MNP} as information carriers and by guiding them via an external magnetic field \changed{in a desired direction}.

For targeted drug delivery and many other biotechnological applications, \acp{MNP} are already widely used~\cite{pankhurst_progress_2009,Zaloga_Development_2014}.
Moreover, \acp{MNP} are also naturally occurring in certain bacteria which use them for navigation in the earth's magnetic field~\cite{bazylinski_magnetosome_2004}.
Synthetic \acp{MNP} usually consist of a \mbox{(bio-)}polymer coating with embedded superparamagnetic iron oxide nanoparticles (SPIONs), which we will refer to as the \emph{magnetic core}, and a non-magnetic coating~\cite{gijs_2004}.
The thin coating ensures biocompatibility and stability, i.e., it prevents agglomeration of the nanoparticles \changed{and can avoid reactions with other molecules}.
\changed{Moreover}, the particle surface can \changed{also} be \emph{functionalized} with binding sites that are selective to specific molecules~\cite{veiseh_design_2010}.
In this way, \acp{MNP} can directly bind to cell membrane receptors or carry target molecules which in turn can be chemically recognized by cells.
Also, by exploiting their magnetic properties, \acp{MNP} themselves can be detected by external devices~\cite{pankhurst_progress_2009}.
This allows for the monitoring of \ac{MNP}-mediated transport processes and also makes \acp{MNP} prime candidates as information carriers as was recently demonstrated in a macroscale \ac{MC} testbed~\cite{Unterweger_Experimental_2018}.
However, most importantly, \acp{MNP} can be externally guided by applying a magnetic field.
Thereby, the magnetic force crucially depends on the magnetic field gradient rather than the magnitude of the magnetic field.
Thus, relatively large forces can be realized by optimizing the design of the magnet to exhibit large magnetic field gradients, see e.g.~\cite{sarwar_optimal_2012} where the arrangement of spatial arrays of permanent magnets was optimized for this purpose in order to overcome blood flow.
Other known use cases of \acp{MNP} include separation of biological material, drug targeting, single cell actuation, and hyperthermia~\cite{pankhurst_applications_2003,pankhurst_progress_2009}.
More specifically, \acp{MNP} can be attached to genes within a cell~\cite{hamad-schifferli_remote_2002}, loaded with drugs~\cite{dobson_magnetic_2006}, attached to cellular ion channels~\cite{dobson_remote_2008} and the cell surface\cite{fakhrullin_cyborg_2012}, and be used to steer magnetotactic bacteria~\cite{martel_flagellated_2009}.
Moreover, \acp{MNP} can be finely controlled such as in magnetic tweezers~\cite{lipfert_quantitative_2009} and also precisely detected for biosensing~\cite{giouroudi_microfluidic_2013}.

Using \acp{MNP} as information carriers for \ac{MC} is motivated by applications such as target detection in blood vessels or within systems of pipelines because they are chemically stable and allow for external control and supervision~\cite{pankhurst_applications_2003}.
As \acp{MNP} are already used in microfluidic settings~\cite{gijs_2004}, in this paper, we analyze \ac{MNP} based \ac{MC} in a simple microfluidic channel to quantify the resulting gains in terms of the amplitude of the impulse response and the symbol error rate.
Although \acp{MNP} are applicable for \ac{MC} in duct channels with arbitrary cross sections, including e.g.\ blood vessels, here we focus on a rectangular cross section for analytical tractability.
\changed[II:1,II:3,II:4]{As one example application of the considered system, we envision the following scenario. In a microfluidic channel, \acp{MNP} might be stored in a reservoir at one of the physical boundaries of the channel. The structured release of the stored \acp{MNP} might be triggered by a certain critical chemical reaction whose occurrence hence can be detected by detecting the particles, e.g., by a susceptibility measurement~\cite{Unterweger_Experimental_2018}. In this manner, an interface from the (microscale) chemical domain to the (macroscale) external domain can be realized. For this example application, the required data rate might not need to be very high but the reliability of the detection of the occurrence of the chemical reaction might be critical. In this paper, we analyze the improvement in communication reliability achievable by employing a magnet for attracting the \acp{MNP} towards the receiver.}

Despite their widespread use in contemporary biotechnology, to the best of our knowledge, for synthetic \ac{MC}, \acp{MNP} have only been considered in~\cite{nakano_externally_2014, kisseleff_magnetic_2017, Unterweger_Experimental_2018}.
In particular, the authors of~\cite{nakano_externally_2014}
\changed[II:3]{remarked}
that \acp{MNP} attached to DNA can
\changed[II:3]{possibly}
initiate gene expression if subjected to an external magnetic field
\changed{but no mathematical analysis was provided.} 
\changed[II:3]{A theoretical model of a}
wearable device detecting changes of inductance when
\changed{a specific concentration of \acp{MNP} passes
through a coil was presented in~\cite{kisseleff_magnetic_2017}.}
Finally,
\changed[II:3,II:6]{as a proof-of-concept of \ac{MNP} based \ac{MC}}
at macroscale, the magnetic properties of \acp{MNP} were used for their detection in
\changed{an experimental flow-based system where diffusion is negligible~\cite{Unterweger_Experimental_2018}.}
However, to the best of the authors' knowledge, the benefits of
\changed{using the magnetic property of \acp{MNP} to guide them with an external magnetic field towards the receiver}
have only been investigated before in the conference version of this work~\cite{wicke_molecular_2017}.

\changed[II:4,II:6]{The theoretical analysis presented in this paper can guide subsequent experimental work. As even for the simple \ac{OOK} modulation scheme, the quantitative impact of a magnetic field on the communication performance is not immediate, a detailed analysis is needed, which is provided in this paper for an example scenario.}

Compared to \cite{wicke_molecular_2017}, in addition to generally extended discussions and more results, the two-dimensional environment with fully reflective boundaries in \cite{wicke_molecular_2017}, is expanded to a three-dimensional environment with partial particle adsorption at the boundaries.
Modeling particle adsorption, which can be due to reactions at the boundaries, is crucial as it can usually not be avoided or might be even desirable in applications relying on these reactions \cite{Nacev_behaviors_2011,Gervais_Mass_2006}.
In addition, we consider a refined particle model where each \ac{MNP} is modeled as a compound embedded with several SPIONs \cite{gijs_2004}.
We also study how a realistic magnetic field impacts the achievable magnetic force close to and far from the magnet, respectively, which was not considered in \cite{wicke_molecular_2017}.

The main contributions of this paper can be summarized as follows:
\begin{enumerate}
    \item
        We propose the use of \acp{MNP} as information carriers and characterize their physical properties.
        Thereby, we model the particle movement in an external magnetic field  as diffusion with drift similar to fluid flow.
        In contrast to fluid flow, a magnetic field can be directed in a desired direction, even towards solid boundaries.
        Moreover, we show that the magnetic force critically depends on the particle size.
    \item
        To illustrate the benefits of using \acp{MNP} for \ac{MC} in a realistic environment, we consider a straight microfluidic channel with rectangular cross section.
        We analyze the time-variant spatial particle distribution subject to the combined effect of fluid flow and magnetic drift and partially adsorbing boundaries~\cite{Gervais_Mass_2006}.
        As it is difficult to produce many particles having exactly the same size, we also take into account the typical log-normal distribution of the particle radius~\cite{kiss_new_1999} in our mathematical expressions.
        We
        \changed{employ}
        the derived impulse response to exemplarily study \ac{OOK} modulation.
    \item
        For the considered model, we calculate the \ac{SER} to evaluate the system performance.
        Thereby, the system is affected by fluid flow which, on the one hand, removes leftover \acp{MNP} from the microfluidic channel but, on the other hand, may also prevent information-carrying \acp{MNP} from reaching the \ac{RX}.
        We show that applying a magnetic force can drastically reduce the \ac{SER} by increasing the number of observed particles.
\end{enumerate}

This paper is organized as follows.
In \cref{sec:system_model}, we introduce the system model and the magnetic properties of the \acp{MNP}.
Based on this model, we derive the exact spatial distribution of the \acp{MNP} and the channel impulse response in \cref{sec:system_analysis}.
In Section~\ref{sec:statistical_analysis}, a statistical analysis of the received signal is provided.
Simulation results are presented in \cref{sec:numerical_results}.
Finally, \cref{sec:conclusion} concludes the paper.

\section{System Model}
\label{sec:system_model}
In this section, we first introduce the considered system geometry.
Then, we discuss general properties of \acp{MNP} and the considered magnetic field.
Finally, we present the adopted modulation and detection schemes.

\subsection{System Geometry}
\begin{figure}[!t]
    \centering
    \subfloat[]{\includegraphics{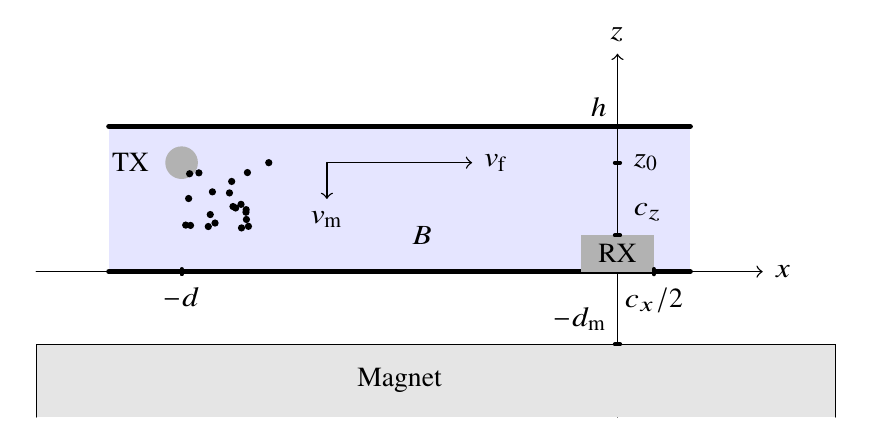}%
    }
    \hfil
    \subfloat[]{\includegraphics{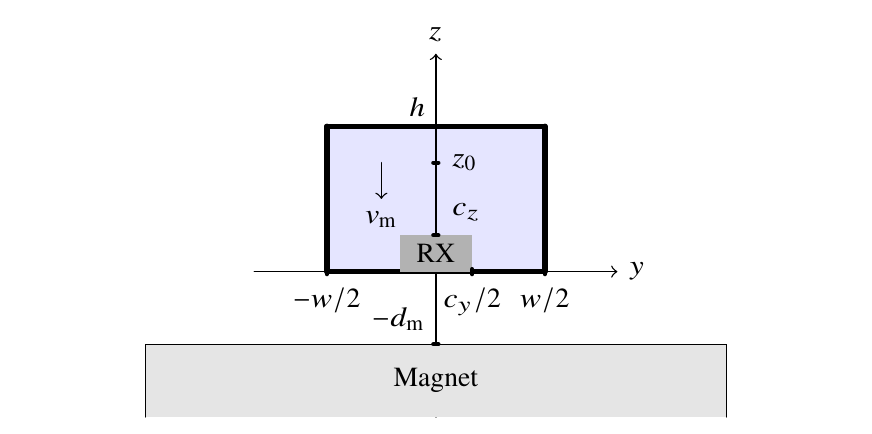}
    }
    \caption{\label{fig:system_model}
        System model geometry in (a) the $x$-$z$-plane and (b) the $y$-$z$-plane.
        \acp{MNP} released by the point \acf{TX} at $x=-d,\;y=0,\;z=z_0$ are schematically shown as black dots and are to be received by the cuboid \acf{RX} of extents $c_x$, $c_y$, and $c_z$ in the $x$-, $y$-, and $z$-coordinates, respectively.
        Fluid flow with velocity $v_\f$ carries the \acp{MNP} downstream past the \ac{RX}.
        A magnet (distance $d_\m$ away from the fluid environment) creating the magnetic field $B$ is dragging the \acp{MNP} downwards towards the \ac{RX} with drift velocity $\vm$.
        \vspace*{-5mm}
    }
\end{figure}

As example model, we consider a straight duct with rectangular cross section%
\footnote{%
    The study of non-rectangular microfluidic system geometries, e.g., a cylindrical geometry as a first order model of a blood vessel, is left for future work, see also \cite{Schaefer_Transfer_2018}.
    For comparison, in Section~\ref{sec:numerical_results}, we present simulation results for both a rectangular and a circular cross section.%
}
with height $h$, width $w$, and infinite axial extent.
In particular, the particles diffuse only within $x\in(-\infty,\infty), z\in[0,h], y\in[-w/2,w/2]$, see \cref{fig:system_model}.
In addition to diffusion, in several applications of mass transport via \acp{MNP}, \acp{MNP} can be lost at boundaries or are required to bind to a target region~\cite{pankhurst_applications_2003}.
Hence, we also model chemical reactions and adhesion of \acp{MNP} on the inner surfaces of the duct by the adsorption coefficient $\adcoeff$ which specifies the rate of particle adsorption over time in the $y$- and $z$-directions with respect to the concentration at the boundary \cite{Gervais_Mass_2006}.
We note that this description naturally includes the special cases of fully reflecting boundaries for $\adcoeff=0$ and fully adsorbing boundaries for $\adcoeff\to\infty$.
For this channel, we assume a point \ac{TX} positioned at $(x,y,z)=(-d,0,z_0)$ wants to deliver a message to the \ac{RX} which is located at $(x,y,z)=(0,0,0)$, see \cref{fig:system_model}.
We also assume that fluid flow with uniform%
\footnote{%
    Flow is a complex phenomenon.
    In general, the flow profile is not uniform, e.g., the flow velocity can be zero at the boundaries and maximal in the center of the duct.
    For simplicity, we will neglect these effects in this paper and focus on uniform flow which still captures the main mass transport phenomena.
    Furthermore, uniform flow is a typical model when the flow is induced by electroosmotic pumps \cite{bruus_2007}.
    Nevertheless, exploring the effect of non-uniform flow on the particle transport in \ac{MC} systems constitutes an interesting topic for future research \cite{wicke_duct_2017}.
}
velocity $v_\f$ carries the \acp{MNP} downstream in positive $x$-direction but possibly past the \ac{RX}.
The \ac{RX} is modeled as a passive transparent observer.
As received signal, we consider the number of particles inside the cuboid defined by $x\in [-c_x/2,c_x/2], y\in [-c_y/2,c_y/2], z\in [0,c_z]$.
To increase the number of particles arriving at the \ac{RX}, a magnet creating the magnetic flux density $B$ is placed below the channel dragging particles in negative $z$-direction%
\footnote{\label{fn:magnet_range}%
\changed[I:7]{In this paper, we assume that the magnet covers the entire space between \ac{TX} and \ac{RX}. This is a realistic assumption when we assume that the transmission range is at microscale while the dimensions of the magnet are at macroscale, e.g., when we consider a microfluidic channel and a handheld magnet. In an application scenario, the magnet could either be specifically provided to improve the communication link or be inherently present such as in a medical screening and be opportunistically exploited for \ac{MC}.}%
}
towards the \ac{RX} with velocity $\vm$.
\changed[II:5]{We note that for $\vm=0$, a conventional molecular communication setup is obtained where the magnetic property of the particles does not play a role.}

In the following subsection, we investigate the dependence of the magnetic drift velocity on the magnetic field.
\begin{figure}[!t]
    \centering
    \vspace*{-5mm}
    \includegraphics{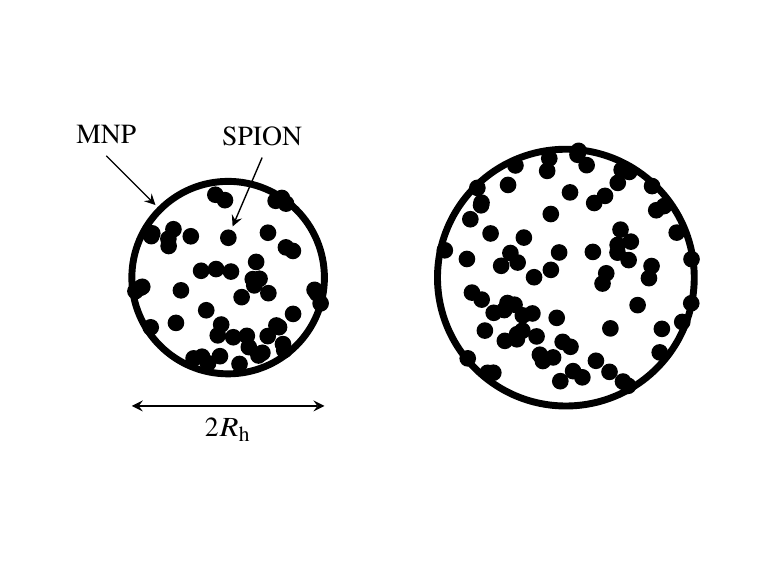}
    \vspace*{-15mm}
    \caption{
        Sketch of two \acp{MNP} consisting of several superparamagnetic iron oxide nanoparticles (SPIONs).
        The two \acp{MNP} have different sizes but the same SPION density $\Cs$.
        \vspace*{-5mm}
    }
    \label{fig:mnp}
\end{figure}
\subsection{Magnetic Nanoparticle Dynamics}
\label{sec:mnp}
Embedding SPIONs in non-magnetic and non-reactive material is a typical approach for designing biocompatible \acp{MNP}~\cite{Lu_Magnetic_2007}.
This yields physical stability of the particles and ensures that undesired reactions with the SPIONs are avoided.
Therefore, we model the \acp{MNP} as composites consisting of $\Ns$ SPIONs embedded in a non-magnetic material, see Fig.~\ref{fig:mnp}.
Thereby, all SPIONs are assumed to have identical volume $\Vs$.
Moreover, we have $\Ns=\Cs\cdot \Vhyd$, where $\Cs$ is the concentration of the SPIONs embedded in one \ac{MNP} and $\Vhyd=4/3\pi\Rhyd^3$ is the volume of an \ac{MNP} of radius $\Rhyd$, see \cref{fig:mnp}.
Because of slight variations in the physical parameters during \ac{MNP} synthesis, the actual particle sizes may differ from the intended size.  
Thereby, the log-normal distribution usually provides a good fit to experimentally observed particle sizes~\cite{kiss_new_1999}.
Motivated by this, for the hydrodynamic particle radius, $\Rhyd$, we assume a log-normal distribution with mean $\mh$ and standard deviation $\sh$.
In the remainder of this paper, we will refer to $\mh$ as the \emph{nominal} particle size, i.e., the target particle size for production of the \acp{MNP}.

We model the externally applied magnetic field by the magnitude of the magnetic flux density $B(z)$, which is a function of $z$ only, i.e., we assume the field to be uniform across $x$ and $y$, reasoning that the gradients in $x$- and $y$-direction are negligible because of the geometry of the magnet.
Thereby, the magnitude of the magnetic flux density gradient, $B'(z)$, increases towards the magnet, i.e., in negative $z$-direction.
Each SPION is affected by the applied magnetic field $B$ in terms of the magnetization $M(B)$.
In particular, considering the thermal energy per particle $\kb \Tf$, where $\kb\approx\SI{1.381e-23}{\meter^2\kilogram\per\second^2\per\kelvin}$ is the Boltzmann constant~{\cite{nelson_biological_2007}} and $\Tf$ the fluid temperature, the average magnetization of a single SPION is dependent on the current value of $B$ and is given by~{\cite[Ch.~4.3.2]{coey_magnetism_2010}}
\begin{equation}
    \label{eq:hysteresis}
    M(B) = \Msat \,{\Langevin}{\left(\frac{\Vs \Msat B}{\kb\Tf}\right)},
\end{equation}
where the \emph{Langevin function} $\Langevin(s)$ is defined as $\Langevin(s)=\coth(s) - 1/s$ and $\coth(s)$ is the hyperbolic cotangent.
Moreover, $\Vs$ is the SPION volume and $M_\mathrm{s}$ denotes the saturation magnetization of the SPION material, i.e., $M(B\to\infty)=M_\mathrm{s}$.
The Langevin function is a point symmetric monotonically increasing function that saturates for small and large arguments to $-1$ and $+1$, respectively.
We note that \eqref{eq:hysteresis} implies $M(B=0)=0$, which is in contrast to larger ferromagnetic materials for which $M$ does not only depend on the current value of $B$ but also on previous values of $B$ and in general is nonzero, i.e., SPIONs show no hysteresis.
This behavior is also referred to as superparamagnetism~\cite{gijs_2004}.

Given the SPION volume $\Vs$ and their average magnetization in \cref{eq:hysteresis}, the force on an \ac{MNP} with $\Ns$ embedded SPIONs%
\footnote{%
    In general, the individual SPIONs may also interact with each other weakening the overall force on the MNPs.
    Eq.~\eqref{eq:magnetic_force} can then be interpreted as an upper bound to the maximum magnetic force.%
}
in a magnetic field $B$ in negative $z$-direction (cf.~\cref{fig:system_model}) is given by~\cite{gijs_2004}
\begin{equation}
    \label{eq:magnetic_force}
    \Fm(z) = -\Ns \Vs \Dz \Big(M\big(B(z)\big)\cdot B(z)\Big),
\end{equation}
where $\Dz$ is the partial derivative with respect to $z$.

The movement of the \acp{MNP} is subject to a magnetic drift with velocity $v_\m$, which is due to the magnetic force $F_\m$, and Brownian motion, which can be characterized by diffusion coefficient~$D$.
It is known that applying a force $F_\m$ on a small particle, such as an \ac{MNP}, immersed in a liquid of viscosity $\eta$ quickly%
\footnote{%
    The time duration for acceleration to the terminal velocity is negligible for low Reynolds numbers \cite{bruus_2007}.
}
accelerates the particle to the terminal velocity $v_\m=F_\m/\zeta$~\cite[Eq.~(4.12)]{nelson_biological_2007}, where $\zeta$ is the friction coefficient which by Stokes' law is given by $\zeta=6\pi\eta R_\hyd$.
In summary, we obtain
\begin{equation}
    \label{eq:myF}
    \vm(z) = -\Rhyd^2\Cs\frac{2\Vs}{9\eta} \Dz \big(M(B(z))\cdot B(z)\big),
\end{equation}
where we have used \changed{$\Ns=\Cs\cdot 4/3\cdot \pi \Rhyd^3$}.
The magnet-induced motion superimposes the motion caused by the fluid flow, i.e., the overall drift vector is given by $(v_\f,0,-v_\m)$, see Fig.~\ref{fig:system_model}.

Eq.~\eqref{eq:myF} can be easily evaluated for a given magnetic field $B(z)$ but does not give simple insight into how the magnetic drift velocity depends on the magnetic field $B(z)$.
To gain insight, the following special cases can be considered.
On the one hand, for very large values of $B$, i.e., when the microfluidic channel is very close to the magnet, according to \eqref{eq:hysteresis}, $M(B)\approx\Msat$ and hence $\vm$ in \eqref{eq:myF} is proportional to $B'(z)$.
On the other hand, for small values of $B$, i.e., when the microfluidic channel is far from the magnet, we can linearize the Langevin function as $\Langevin(s)=\frac{1}{3}s$ which yields $M(B)= \alpha B$, where $\alpha=\Msat^2\Vs/(3\kb\Tf)$.
In this case, $\vm$ is proportional to $\Dz\big(B(z)\big)^2=2B(z) B'(z)$.
In summary, for small and large $B$, $\vm$ in \eqref{eq:myF} can be approximated by
\begin{subnumcases}{\label{eq:vmspecial}\vm(z)=}
    -\Rhyd^2\Cs\frac{4\Vs}{9\eta} \chi B(z) B'(z), & $B$ small
    \label{eq:vmBsmall}
    \\ 
    -\Rhyd^2\Cs\frac{2\Vs}{9\eta} \Msat B'(z), & $B$ large.
    \label{eq:vmBlarge}
\end{subnumcases}
Thereby, the force points towards the magnet because this is the direction of increasing magnetic field strength.
From \eqref{eq:myF} and \eqref{eq:vmspecial}, we can conclude that the magnetic drift velocity crucially depends on the magnetic field gradient.

By thermodynamic reasoning~\cite[Ch.~4]{nelson_biological_2007}, friction coefficient $\zeta$ is linked to the diffusion coefficient $D$ by the Einstein relation $\kb\Tf = D\zeta$.
Hence, given the viscosity $\eta$ and the temperature of the fluid $\Tf$, $D$ can be determined as
\begin{equation}
    \label{eq:myD}
    D = \frac{\kb \Tf}{6\pi\eta \Rhyd},
\end{equation}
which also depends on the \ac{MNP} size but to a lesser degree than $\vm$ in \eqref{eq:myF}.
In contrast, the fluid flow velocity $\vf$ is not affected by (small) changes of $\Rhyd$.

\subsection{Magnetic Field}
Eq.~\eqref{eq:myF} is valid for general magnetic fields $B(z)$.
As one particular example, we consider the magnetic field created by a cylindrical magnet of length $\Lmag$ and radius $\Rmag$ along its symmetry axis~\cite{furlani_nanoscale_2008}
\begin{equation}
    \label{eq:Bz}
    B(z-d_\m) = \frac{B_0}{2} \left(\frac{z + \Lmag}{\sqrt{(z+\Lmag)^2 + \Rmag^2}} - 
    \frac{z}{\sqrt{z^2 + \Rmag^2}}\right),
\end{equation}
where $B_0$ is a system parameter reflecting the strength of the magnet.
Furthermore, the derivative of $B(z)$, is obtained as
\begin{equation}
    \label{eq:dBz}
    B'(z-d_\m) = \frac{B_0}{2} \left(\frac{\Rmag^2}{\big((z+\Lmag)^2+\Rmag^2\big)^{3/2}} - 
    \frac{\Rmag^2}{\big(z^2+\Rmag^2\big)^{3/2}}\right).
\end{equation}
Eqs.~\eqref{eq:Bz} and \eqref{eq:dBz} are sufficient for describing the magnetic field within the assumed microfluidic channel where $\Rmag\gg d,w$.

Because the height $h$ of the considered microfluidic channel is assumed to be much smaller than the length $\Lmag$ of the magnet we can safely assume that both $B(z)$ and $B'(z)$ are constant for $z\in[0,h]$.
As a consequence, $\vm(z)$ in \eqref{eq:myF} can also be assumed to be constant within the considered environment.
Hence, in the remainder of this paper, we will assume $\vm(z)$ to be constant for $z\in[0,h]$ and simply denote it by $\vm$ \cite{Nacev_behaviors_2011}.

\subsection{Modulation and Detection}
\label{sec:channelModel}
%
To illustrate the principal benefits of using \acp{MNP}, we focus on the following basic \ac{MC} model~\cite{farsad_comprehensive_2016}.
Binary information symbols, $b[i]$, are modulated by \ac{OOK}.
Assuming instantaneous particle release, for transmitting $b[i]=1$ and $b[i]=0$, the point source \ac{TX} instantaneously releases $N_\TX$ and 0 \acp{MNP} at the beginning of each time slot of length $T$, respectively.
We assume that the \ac{RX} is perfectly synchronized with the \ac{TX}, i.e., the \ac{RX} knows the symbol interval $T$ and when transmission starts and ends \cite{Jamali_Symbol_2017}.
By counting the number of particles within its volume, the \ac{RX} takes samples at times $iT+t_0$, where $t_0$ is a time offset after which, for each symbol interval $i$, particles can be expected within the receiver volume.
For detection, in each symbol interval, the number of counted particles at the \ac{RX}, $n_\RX[i]$, is compared to a threshold $\xi$, i.e., the detected symbols are given by
\begin{equation}
    \label{eq:bhat}
    \hat{b}[i] = 
    \begin{cases}
        0,	&	n_\RX[i] < \xi \\
        1,	&	n_\RX[i] \geq \xi.
    \end{cases}
\end{equation}
%

\section{Performance Analysis for Microfluidic Channel}
\label{sec:system_analysis}
In this section, we derive an analytical expression for the time-variant spatial \ac{MNP} distribution for a single particle by solving the diffusion equation with drift for the system in \cref{fig:system_model}.
Then, equipped with the spatial \ac{PDF} $p(x,y,z;t)$ (the so called Green's function in case of an instantaneous point release), we calculate the probability of observing a single particle of size $\Rhyd$ within the \ac{RX} volume.
In this section, we only consider a single fixed particle size $\Rhyd$.

\subsection{Green's Function}
For the environment depicted in \cref{fig:system_model}, the particle movement in the $x$-, $y$-, and $z$-directions is uncoupled and hence the time-varying \ac{PDF} of the position of an \ac{MNP} can be written as $p(x,y,z;t) = p_x(x;t) p_y(y;t) p_z(z;t)$.
Thereby, the axial distribution $p_x(x;t)$ corresponds to an unbounded environment with constant drift $v_\f$.
Hence, this distribution is readily obtained as~\cite[Eq.~(4.39)]{schulten_lectures_2000}
\begin{equation}
    \label{eq:px}
    p_x(x;t) = \frac{1}{\sqrt{4D\pi t}} {\exp}{\left(\frac{-(x+d-v_\f t)^2}{4Dt}\right)},
\end{equation}
where the mean particle $x$-coordinate, $\meanx(t)=-d+v_\f t$, arrives at the center of  the \ac{RX} at time $t_1=d/v_\f$.
Determining the vertical distribution $p_z(z;t)$ is more challenging because of the combination of a bounded environment and particle drift.
On the other hand, once $p_z(z;t)$ is known, we can easily obtain $p_y(y;t)$ as a special case where we account for the absence of magnetic drift in $y$-direction and employ a shifted coordinate system $y\in[-w/2,w/2]$ (see Fig.~\ref{fig:system_model}).

Therefore, in the following, we first calculate $p_z(z;t)$.
In particular, we consider the underlying \ac{PDE}, which is the diffusion equation with drift with appropriate boundary conditions.
The diffusion equation with drift can be expressed as $\D{t}p_z=-\D{z}J_z$ where $J_z$ is a quantity referred to as \emph{probability flux} (unit: \si{\per\second}) \cite{schulten_lectures_2000}.
This probability flux consists of two components, one representing particle diffusion and one representing particle movement due to the magnetic force:
\begin{equation}
    J_z(z;t) = -D\D{z} p_z(z;t) - v_\m p_z(z;t).
\end{equation}
Thereby, the adsorbing boundary conditions are specified by~\cite[Eq.~(4.26)]{schulten_lectures_2000}
\begin{equation}
    \label{eq:boundary_flux}
    J_z(z;t)=\adcoeff p_z(z;t),
\end{equation}
for $z=0,h$, and $t>0$.
Moreover, the initial particle $z$-position is given by $z=z_0$ of the point source \ac{TX}.
In summary, $p_z(z;t)$ for $0<z<h$ and $t>0$ is obtained by solving the following \ac{PDE} with boundary and initial conditions:
\begin{IEEEeqnarray}{rCll}
    \label{eq:pde_bounded}
    \D{t} p_z(z;t) & = &v_\m\D{z} p_z(z;t) + D\D[2]{z} p_z(z;t), & \IEEEyesnumber\IEEEyessubnumber* \\
    \frac{\partial}{\partial z} p_z(z;t)     &= &\frac{1}{D} (\adcoeff - v_\m) p_z(z;t), & z=0 \\
    \frac{\partial}{\partial z} p_z(z;t)     &= &-\frac{1}{D}(\adcoeff + v_\m)p_z(z;t), & z=h \\
    p_z(z;t) &= &\delta(z-z_0), & t = 0.
\end{IEEEeqnarray}
%
The solution to \eqref{eq:pde_bounded}
\changed[II:5]{cannot be directly obtained, e.g., from \cite{carslaw_conduction_1986}, and requires a careful derivation. The result is presented in the following theorem.}
To this end, the following auxiliary variables are defined:
\begin{equation}
    u = \frac{\vm}{2D} \qquad \text{and} \qquad \mcoeff = \frac{\adcoeff}{D}.
\end{equation}

\begin{theorem}
    \label{theorem:main}
The \ac{PDF} $p_z(z;t)$ can be expressed in form of an infinite series as
\begin{IEEEeqnarray}{c}
    \label{eq:constant_bounded_exact_pdf}
    p_z(z; t) =  {\exp}{\left(-u(z-z_0)- Du^2t\right)}
               \sum_{n=0}^N a_n\exp(-Ds_n^2 t)Z_n(z),\nonumber\\*
\end{IEEEeqnarray}
where $s_n,\; n=0,1,\dots$, are referred to as \emph{eigenvalues}, $Z_n(z)$ as \emph{eigenfunctions}, $a_n,\; n=0,1,\ldots$ are series coefficients, and $N$ is the number of series coefficients to be considered for numerical evaluation.

The eigenfunctions can be expressed as
\begin{equation}
    \label{eq:cossol}
    Z_n(z) = \cos(s_n z) + \frac{\mcoeff-u}{s_n} \sin(s_n z),
\end{equation}
and the eigenvalues $s_n$ are the non-negative real or imaginary solutions to
\begin{equation}
    \label{eq:taneq}
    \tan(s_nh) = \frac{2s_n\mcoeff}{s_n^2 - \mcoeff^2 + u^2}.
\end{equation}

The coefficients $a_n$ in~\eqref{eq:constant_bounded_exact_pdf} can be determined as
\begin{equation}
    \label{eq:an}
    a_n = \frac{Z_n(z_0)}{\lVert Z_n \rVert^2}
\end{equation}
where $\norm{Z_n}^2 = \int_0^h |Z_n(z)|^2\diff z$ is given by
\begin{IEEEeqnarray}{rCll}
    \label{eq:normZn}
    \norm{Z_n}^2 &= &\frac{1}{4s_n^3} \big( &2s_n[h(s_n^2 + \alpha^2) -\alpha] + 2s_n\alpha\cos(2hs_n)\nonumber \\
                 & & & +(s_n^2-\alpha^2) \sin(2hs_n)\big)
\end{IEEEeqnarray}
with $\alpha=u-\mcoeff$.
\end{theorem}
\begin{IEEEproof}
    Please refer to the appendix.
\end{IEEEproof}

In the following, we discuss some properties of the solution given in Theorem~\ref{theorem:main} before deriving the spatial distribution for the $y$-coordinate.
It can be shown that, $s_n,\; n=1,2,\dots$, the positive real-valued solutions to \eqref{eq:taneq}, lie within $[\pi/h, \infty)$.
However, there exists a critical magnetic drift velocity parameter, $\ucrit$, and $s_0$ is real, imaginary, and diminishes to zero when $u<u_\crit$, $u>u_\crit$, and $u=u_\crit$, respectively.
This critical drift velocity parameter is given by
\begin{equation}
    \label{eq:ucrit}
    \ucrit = \sqrt{\frac{2}{h}\mcoeff + \mcoeff^2}.
\end{equation}
On the one hand, for $u<\ucrit$, $Z_0(z)$ and $s_0$ are directly given by \eqref{eq:cossol} and by solving \eqref{eq:taneq}, respectively.
On the other hand, for $u\geq\ucrit$ some simplifications are possible.
Hence, when evaluating \eqref{eq:cossol} for $u<\ucrit$, $u>\ucrit$ (i.e., $s_0=\im\sigma$, where $\im=\sqrt{-1}$), and $u=\ucrit$ (i.e., $s_0\to0$), we obtain
\begin{subnumcases}{\hspace*{-1cm}Z_0(z)=\label{eq:z0sol}}
    \text{Eq.~\eqref{eq:cossol} with $n=0$}, & $u<\ucrit$ \\
    \cosh(\sigma z) + \frac{\mcoeff - u}{\sigma} \sinh(\sigma z), & $u>\ucrit$ \label{eq:coshsol0}\\
    1 + (\mcoeff-u)\cdot z, & $u=\ucrit$.
    \label{eq:affsol0}
\end{subnumcases}
Moreover, for \eqref{eq:coshsol0}, from \eqref{eq:taneq}, $\sigma$ can be found by solving

\begin{equation}
    \label{eq:tanheq}
    \tanh(\sigma h) = \frac{-2\sigma\mcoeff}{\sigma^2 + (\mcoeff^2 - u^2)}.
\end{equation}
Thereby, for the first eigenvalue $s_0$, from \eqref{eq:taneq} and \eqref{eq:tanheq} we obtain the following bounds
\begin{equation}
    \label{eq:s0_bounds}
    -u^2 \leq s_0^2 < \frac{\pi}{h}.
\end{equation}
For completeness, for $n=0$, from \eqref{eq:normZn} we obtain
\begin{subnumcases}{\hspace*{-0.5cm}\norm{Z_0}^2=}
    \label{eq:z0norm1}
    \text{Eq.~\eqref{eq:normZn} with $n=0$}, & $u<\ucrit$ \\
    \begin{IEEEeqnarraybox}{ll}
        \frac{-1}{4\sigma^3}\big\{&2\sigma[(\alpha^2-\sigma^2)h-\alpha]\\
         &+ 2\sigma\alpha\cosh(2\sigma h) \\
         &- (\sigma^2+\alpha^2)\sinh(2\sigma h)\big\}
    \end{IEEEeqnarraybox} & $u>\ucrit$
    \\
    (1-\alpha)h + \frac{\alpha^2}{3}h^3 & $u=\ucrit$.
\end{subnumcases}

Now that we have obtained $p_z(z;t)$, we can straightforwardly obtain $p_y(y;t)$ from $p_z(z;t)$ by substituting $h$ with $w$, $z$ with $y+w/2$, $z_0$ with $y_0+w/2$, $s_n$ with $\sy_n$, and $u$ with $0$ as follows
\begin{equation}
    \label{eq:py}
    p_y(y;t) = \sum_{n=0}^N b_n \exp(-D\sy_n^2) Y_n(y),
\end{equation}
where from \eqref{eq:cossol}
\begin{equation}
    \label{eq:Yncossol}
    Y_n(y) = \cos(\sy_n(y+w/2)) + \frac{\kappa}{\sy_n} \sin(\sy_n(y+w/2)),
\end{equation}
and from \eqref{eq:taneq}, $\sy_n$ is the solution to
\begin{equation}
    \label{eq:taneqy}
    \tan(\sy_n w) = \frac{2 \sy_n \mcoeff}{\sy_n^2 - \mcoeff^2}.
\end{equation}
Moreover, from \eqref{eq:an}
\begin{equation}
    \label{eq:bn}
    b_n = \frac{Y_n(y_0)}{\norm{Y_n}},
\end{equation}
where similar to \eqref{eq:normZn}
\begin{IEEEeqnarray}{rCll}
    \label{eq:Ynnorm}
    \norm{Y_n}^2 &=& \frac{1}{4\sy_n^3} \Big(&2\sy_n[w(\sy_n^2 +\mcoeff^2) + \mcoeff] - 2\sy_n\mcoeff\cos(2w\sy_n) \nonumber\\
                 & & &+ (\sy_n^2 - \mcoeff^2)\sin(2w\sy_n)\Big).
\end{IEEEeqnarray}
For completeness, if $\mcoeff=0$ where $\sy_0\to0$, \eqref{eq:Yncossol} simplifies to $Y_0(y)=1$ and \eqref{eq:bn} becomes $b_0=1/w$ because, in this case, \eqref{eq:Ynnorm} yields $\norm{Y_0}^2=w$.

Now, $p(x,y,z;t)$ can be determined by combining \eqref{eq:px}, \eqref{eq:constant_bounded_exact_pdf}, and \eqref{eq:py}.

\subsection{Probability of Particle Observation}
Using the \ac{PDF} $p(x,y,z;t)$, we can now obtain the probability of observing a particle within the \ac{RX} volume, $P_\ob(t)$, as
\begin{IEEEeqnarray}{rCl}
    \label{eq:received_in_tube}
    P_\ob(t) &= &\int_{-c_x/2}^{c_x/2} \int_{-c_y/2}^{c_y/2} \int_0^{c_z} p(x,y,z;t)\de z\de y\de x \nonumber\\
             &= &P_{\ob,x}(t) P_{\ob,y}(t) P_{\ob,z}(t),
\end{IEEEeqnarray}
where $P_{\ob,x}(t)$, $\Poby(t)$, and $P_{\ob,z}(t)$ are the probabilities of observing a particle within the \ac{RX} $x$-, $y$-, and $z$-coordinates $[-c_x/2,c_x/2]$, $[-c_y/2,c_y/2]$, and $[0, c_z]$, respectively.
Integrating \cref{eq:px} from $-c_x/2$ to $c_x/2$ yields
\begin{equation}
    \label{eq:one_dimension_observation_probability}
    P_{\ob,x}(t) = \frac{1}{2} \left[{\erf}{\left(\frac{\meanx + \frac{1}{2}c_x}{\sqrt{4Dt}}\right)} - {\erf}{\left(\frac{\meanx - \frac{1}{2}c_x}{\sqrt{4Dt}}\right)}\right],
\end{equation}
where $\erf(\cdot)$ is the error function.

Furthermore, integrating \cref{eq:constant_bounded_exact_pdf} from $0$ to $c_z$, we obtain $P_{\ob,z}(t)$ as
\begin{equation}
    \label{eq:constant_bounded_exact}
    P_{\ob,z}(t) = {\exp}{\big(-u^2 Dt\big)} \sum_{n=0}^N a_n {\exp}{\bigg(-Ds_n^2 t\bigg)}\Zint_n,
\end{equation}
where the coefficients $\Zint_n$ are defined as
\begin{equation}
    \label{eq:Zint}
    \Zint_n = \int_0^{c_z} {\exp}{\left(-u(z-z_0)\right)}\cdot Z_n(z)\diff z.
\end{equation}
For $Z_n(z)$ in \eqref{eq:cossol} these coefficients can be obtained as
\begin{IEEEeqnarray}{ll}
    \label{eq:Zintn}
    \Zint_n = &\frac{\exp(u(z_0-c_z))}{s_n(s_n^2+u^2)} \cdot \bigg[(s_n^2 +u^2-\mcoeff u)\sin(c_zs_n) \nonumber\\
            &\qquad- \mcoeff s_n\cos(c_zs_n) + \mcoeff s_n\exp(c_zu)\bigg]
\end{IEEEeqnarray}
which is directly applicable for $n>0$.
On the other hand, for $n=0$, from \eqref{eq:z0sol} and \eqref{eq:Zint}, we obtain $\Zint_0=\exp(uz_0-c_zu)\cdot\Zint_0'$ with
\begin{subnumcases}{\hspace*{-5mm}\Zint_0'=\label{eq:Zint0}}
    \text{Eq.~\eqref{eq:Zintn} with $n=0$}, & $u<\ucrit$ \\
    \begin{IEEEeqnarraybox}{ll}
        \Big[&(u^2-\sigma^2-\mcoeff u)\sinh(c_z\sigma)\\
             &- \mcoeff\sigma\cosh(c_z\sigma) \\
             &+\mcoeff\sigma\exp(c_zu)\Big]/\big[\sigma(u^2-\sigma^2)\big]
    \end{IEEEeqnarraybox} & $u>\ucrit$\\
    \frac{\mcoeff\exp(c_zu) -\mcoeff +c_zu^2 -c_z\mcoeff u}{u^2} & $u=\ucrit$.
\end{subnumcases}

Now, after obtaining $\Pobx(t)$ in \eqref{eq:one_dimension_observation_probability} and $\Pobz(t)$ in \eqref{eq:constant_bounded_exact}, we determine $\Poby(t)$.
Integrating $p_y(y;t)$ in \eqref{eq:py} from $y=-c_y/2$ to $y=c_y/2$ yields
\begin{equation}
    \label{eq:Poby}
    \Poby(t) = \sum_{n=0}^N b_n \exp(-D \sy_n^2 t) \Yint_n,
\end{equation}
where $\Yint_n=\int_{-c_y/2}^{c_y/2} Y_n(y)\de y$.
For $n\geq 0$ and $\kappa>0$, from \eqref{eq:Yncossol} we obtain
\begin{equation}
    \Yint_n = \frac{2}{\sy_n}{\sin}{\left(\frac{\sy_n}{2}c_y\right)} \left[ {\cos}{\left(\frac{\sy_n}{2}w\right)} + \frac{\kappa}{\sy_n^2} {\sin}{\left(\frac{\sy_n}{2}w\right)}\right].
\end{equation}
On the other hand, for $\kappa=0$ where $Y_0(y)=1$, we obtain
\begin{equation}
    \Yint_0= c_y.
\end{equation}

In the following, we will consider several special cases to get more insight into the general solution for the particle observation probabilities in \eqref{eq:constant_bounded_exact} and \eqref{eq:Poby}.

\subsubsection{Asymptotic behavior}
Solution \eqref{eq:constant_bounded_exact} can be simplified for $t\gg h^2/D$, where all terms $n>0$ can be neglected, as these terms are associated with increasingly large eigenvalues $s_n$, which by \eqref{eq:constant_bounded_exact} contribute less and less to the overall solution.
Then, by setting $N=0$ in \eqref{eq:constant_bounded_exact}, we obtain the asymptotic quasi-steady state solution as follows
\begin{equation}
    \label{eq:Pobzeq}
    \Pobzeq(t) = \frac{Z_0(z_0)}{\norm{Z_0}^2} \cdot \exp(-D(u^2+s_0^2)t)\cdot \Zint_0,
\end{equation}
where $Z_0(z)$ is given in \eqref{eq:z0sol} and $\Zint_0$ is given via \eqref{eq:Zint0}.
We note that by \eqref{eq:s0_bounds}, $\Pobzeq(t)$ decays exponentially over time with time constant $D(u^2+s_0^2)$, unless $\mcoeff=0$ (i.e., there is no adsorption), in which case $s_0^2=-u^2$ and hence $\Pobzeq(t)$ is constant (see discussion below).

In a similar manner, for $N=0$ in \eqref{eq:Poby}, we obtain
\begin{equation}
    \Pobyeq(t) = \frac{Y_0(y_0)}{\norm{Y_0}^2} \exp(-D\sy_0^2 t) \cdot \Yint_0,
\end{equation}
which is valid for $t\gg w^2/D$.

In summary, for $t\gg\max\{w,h\}^2/D$, the observation probability can be simplified to

\begin{equation}
\label{eq:Pob_eq_approx}
\Pobeq(t) = \Pobx(t) \Pobyeq(t) \Pobzeq(t).
\end{equation}
We note that for $\Pobx(t)$ we do not attempt to find an asymptotic expression since the overall behavior is already given by the simple term in \eqref{eq:one_dimension_observation_probability}.

\subsubsection{Fully reflecting boundaries}
For fully reflecting boundaries where $\mcoeff=0$, by \eqref{eq:ucrit}, we have $\ucrit=0$, and hence for $u>0$ and $u=0$, $Z_0(z)$ is obtained from \eqref{eq:coshsol0} and \eqref{eq:affsol0} with $s_0=\im u$ and $s_0=0$, respectively.
Moreover, from \eqref{eq:taneq}, $s_n=n\pi/h,\;n=1,2,\dots$.

Hence, from \eqref{eq:z0sol}, we obtain
\begin{IEEEeqnarray}{rCl}
    \Pobz(t) &= &\frac{1-\exp(-2c_z u)}{1-\exp(-2uh)} \nonumber\\
             & &+ \exp(-u^2 Dt) \sum_{n=1}^N a_n \exp(-Ds_n^2 t) \cdot \Zint_n,
\end{IEEEeqnarray}
where
\begin{equation}
    a_n = \frac{2}{h} \frac{\cos(s_nz_0) - \frac{u}{s_n}\sin(s_nz_0)}{1+\frac{u^2}{s_n^2}}
\end{equation}
and
\begin{equation}
    \Zint_n = \exp(u(z_0-u)) \frac{\sin(c_zs_n)}{s_n}.
\end{equation}
In this way, we obtain the solution in~\cite[Eq.~(17)]{wicke_molecular_2017} which is characterized by a non-zero time-independent steady state
\begin{equation}
    \Pobzeq(t) = \frac{1-\exp(-2c_z u)}{1-\exp(-2uh)},
\end{equation}
which approaches $\Pobzeq=1$ for $u\to\infty$, i.e., if the magnetic force is very large, as all particles settle at $z=0$.

In a similar manner, we obtain
\begin{equation}
    \Poby(t) = \frac{c_y}{w} + \sum_{n=1}^N b_n \exp(-D\sy_n^2 t) \Yint_n,
\end{equation}
where
\begin{equation}
    b_n = \frac{2}{w} {\cos}{\left(\sy_n(y_0+\frac{w}{2})\right)}
\end{equation}
and
\begin{equation}
    \Yint_n = \frac{2}{\sy_n} {\cos}{\left(\frac{\sy_n}{2}w\right)}\cdot {\sin}{\left(\frac{\sy_n}{2}c_y\right)}.
\end{equation}
Clearly, $\Pobyeq(t)=c_y/w$ for $t\to\infty$, as in this case, the particles are uniformly distributed with respect to $y$.
Moreover, for $c_y=w$, $\Poby(t)=1$ because no particles are lost by adsorption at the boundaries.

\subsubsection{Fully adsorbing boundaries}
For fully adsorbing boundaries where $\mcoeff\to\infty$, by \eqref{eq:ucrit} $\ucrit\to\infty$ and hence for any finite $u$, we have $u<u_\crit$.
As a consequence, $Z_0(z)$ is directly given by \eqref{eq:cossol} for $n=0$.
Furthermore, in this case, \eqref{eq:taneq} reduces to $\tan(s_n h)=0$, and we obtain $s_n=(n+1)\pi/h,\;n=0,1,\dots$.
Then, we can derive
\begin{IEEEeqnarray}{rCl}
    \label{eq:Pobzfullyad}
    \Pobz(t) &= &\exp(-u^2Dt) \nonumber \\
             &&
    \begin{IEEEeqnarraybox}{ll}
         \cdot\sum_{n=0}^N&\frac{2s_n}{h} \sin(s_nz_0) \exp(-Ds_n^2t)\cdot \exp(u(z_0-c_z))\\
                          &\cdot\frac{\exp(c_zu) - \cos(c_zs_n) - \frac{u}{s_n}\sin(c_zs_n)}{s_n^2 +u^2}.
    \end{IEEEeqnarraybox}
\end{IEEEeqnarray}
For $z_0=0$ and $z_0=h$, $\Pobz(t)=0$ as particles get immediately adsorbed where they are released.
Moreover, $\Poby(t)$ is obtained in a similar manner by substituting $u$ with $0$, $z_0$ with $y_0+w/2$, and $s_n$ with $\sy_n$ in \eqref{eq:Pobzfullyad}.

\section{Statistical Analysis}
\label{sec:statistical_analysis}
In this section, we discuss the statistical properties of the considered system.
The number of observed particles is random due to the log-normal particle size distribution of $\Rhyd$ and the random arrival of the particles at the \ac{RX} due to diffusion.
We derive the expected number of observed particles, which is a function of time and will be referred to as (channel) \emph{impulse response}.
Furthermore, given the impulse response, we determine the average received signal and the \ac{SER}.

\subsection{Particle Statistics}
By \eqref{eq:myF}, the magnetic drift velocity can be written as
\begin{equation}
    \label{eq:vmlog}
    \vm = \vmnom \left(\frac{\Rhyd}{\mh}\right)^2,
\end{equation}
where $\Rhyd$ is log-normal distributed with mean $\mh$ corresponding to the magnetic drift velocity obtained for the nominal particle size, which is denoted by $v_{\m,0}$.
By \eqref{eq:vmlog}, $\vm$ is also log-normal distributed~\cite{Papoulis_Probability_2002}.

Similarly, by \eqref{eq:myD} the diffusion coefficient is also a log-normal distributed random variable and can be written as
\begin{equation}
    \label{eq:Dlog}
    D = D_0 \frac{\mh}{\Rhyd},
\end{equation}
where $D_0$ denotes the diffusion coefficient for nominal particle sizes.
On the other hand, the flow velocity $\vf$ is unaffected by the value of $\Rhyd$.

\subsection{Impulse Response}
In principle, the mean particle observation probability is given by the expected value of $\Pob(t)$ with respect to the log-normal distributed particle radius $\Rhyd$ as
\begin{equation}
    \label{eq:meanPob}
    \E\{\Pob(t)\} = \int_0^\infty \Pob(t;r)\cdot f_{\Rhyd}(r)\de r,
\end{equation}
where $\E\{\cdot\}$ denotes expectation and $f_{\Rhyd}(r)$ is the log-normal \ac{PDF} of the particle radius with mean $\mh$ and standard deviation $\sh$.
In \eqref{eq:meanPob}, $\vm$, $D$, and also $s_n$ and $\sy_n$ via \eqref{eq:taneq} and \eqref{eq:taneqy} depend on $\Rhyd$.
Thereby, we will refer to the mean number of observed particles $\Nobmean(t)=\Ntx\cdot\E\{\Pob(t)\}$ as channel impulse response.
We can numerically evaluate \eqref{eq:meanPob} for $\adcoeff=0$ and $\adcoeff\to\infty$ where $s_n and \sy_n,\; n=1,2,\dots$, are known analytically.
However, for general $0<\adcoeff<\infty$, solving the above integral, even numerically, is difficult because $s_n$ can be obtained only indirectly by solving the fixed-point equation \eqref{eq:taneq}.
As a remedy, we employ a Monte Carlo integration as follows:
\begin{equation}
    \label{eq:myNob}
    \Nobmean(t) = \sum_{i=1}^{N_\TX} P_{\ob,i}(t),
\end{equation}
where $P_{\ob,i}(t)$ is the probability of observing particle $i$ at time $t$ which is a random variable due to the log-normal distributed particle size $\Rhyd$.
Thereby, for $\sh>0$, the $P_{\ob,i}(t)$, $i=1,2,\dots,N_\TX$, depend on the particle sizes.
In particular, in \eqref{eq:one_dimension_observation_probability}, \eqref{eq:Poby}, and \eqref{eq:constant_bounded_exact}, $v_\m$ and $D$ depend on $\Rhyd$ via \cref{eq:vmlog,eq:Dlog}, respectively.
We note that for $\Ntx\to\infty$, $\Nobmean(t)/\Ntx$ in \eqref{eq:myNob} will approach \eqref{eq:meanPob} by the law of large numbers \cite{Papoulis_Probability_2002}.
On the other hand, for $\sh=0$, we have $P_{\ob,i}(t)=P_{\ob}(t),\;\forall i$, i.e., all particles have the same (nominal) particle size $\mh$.
In this case, $\Nobmean(t) = N_\TX P_{\ob}(t)$ which we will refer to as the \emph{nominal impulse response}.

\subsection{Symbol Error Rate}
Equipped with the impulse response derived in the previous subsection, we now aim to determine the symbol error rate for \ac{OOK} modulation with $\Ntx$ particles following the described log-normal size distribution.

Using~\cite[Eq.~(30)]{noel_improving_2014}, the average number of observed particles $\nRXmean[i]$ in the $i$-th time slot due to the transmitted symbols $b[i]\in\{0, 1\},\; 0\leq i<K$, is given by
\begin{equation}
    \nRXmean[i] = \sum_{j=0}^{i} b[j]\Nobmean((i-j)T + t_0),
\end{equation}
where $\Nobmean(t)$ is given in \eqref{eq:myNob}.
Assuming that the only impairments are the random number of observed particles due to their diffusive arrivals and \ac{ISI} caused by previously released molecules, we can define the \emph{average} \ac{SER} over all $K$ sent symbols as
\begin{equation}
    \label{eq:myPeDef}
    P_\mathrm{e} = \frac{1}{2^K}\sum_{b\in\mathbb{B}(K)} \left[\frac{1}{K}\sum_{i=0}^{K-1} {\Pr}{\left(\hat{b}[i]\neq b[i];\; b[j\leq i]\right)} \right],
\end{equation}
where $\mathbb{B}(K)$ is the set of all $2^K$ possible binary sequences of length $K$ and $b[j\leq i]$ are the transmitted symbols for time slots $j=0,1,\dots,i$.
In other words, the expression ${\Pr}{\left(\hat{b}[i]\neq b[i];\; b[j\leq i]\right)}$ represents the probability of detecting $b[i]$ incorrectly given the transmitted sequence $b[j]$ for $0\leq j\leq i$.

In general, using the decision rule in \eqref{eq:bhat}, the probability of making an error for the $i$-th symbol can be written as
\begin{equation}
    \label{eq:myPe}
    \Pr(\hat{b}[i]\neq b[i];\; 
    b[j\leq i]) =
    \begin{cases}
        p_\xi(b[j\leq i]), & b[i] = 1 \\
        1-p_\xi(b[j\leq i]), & b[i] = 0,
    \end{cases}
\end{equation}
where%
\footnote{
    We note that in \eqref{eq:myPe}, the probability of observing less than $\xi$ \acp{MNP} at the $i$-th sampling time, given $b[j]$ for $0\leq j\leq i$, $p_\xi(b[j\leq i])$ implicitly depends on the current symbol $b[i]$ and hence the sum of $p_\xi(b[j\leq i])$ (when $b[i]=1$) and $1-p_\xi(b[j\leq i])$ (when $b[i]=0$) do not add up to 1.
}
$p_\xi(b[j\leq i])=\Pr(n_\RX[i]< \xi;\; b[j\leq i])$.
Here, $n_\RX[i]$ is the sum of $\Ntx$ Bernoulli random variables with different success probabilities $\Pob(t_0+kT)$ due to the log-normal distributed particle sizes.
Despite these complications, similar to \cite{noel_improving_2014}, $n_\RX[i]$ can be well approximated%
\footnote{%
    The accuracy of this approximation with regard to the derived performance metrics is verified via simulation in Section~\ref{sec:numerical_results}.
}
by a Poisson random variable with mean $\nRXmean[i]$, see also~\cite{le_cam_approximation_1960}.
In this case, $p_\xi=\Pr(n_\RX[i]\leq \xi-1;\; b[j\leq i])$ is the Poisson \emph{cumulative distribution function} $\PoissonCDF(\xi-1;\nRXmean[i])$ with mean $\nRXmean[i]$ evaluated at $\xi-1$ \cite{Papoulis_Probability_2002}.

Because the constant fluid flow washes particles away from the \ac{RX}, for sufficiently large modulation intervals $T$, no superfluous particles from previous transmissions remain at the \ac{RX}.
In this case, if there is no \ac{ISI}, then for any $\xi\geq 1$, $b[i]=0$ is always detected correctly because in this case $\nRXmean[i]=0$. 
On the other hand, if $b[i]=1$, then $\nRXmean[i]=\Nobmean(t_0)$ and the optimal detection threshold is $\xi=1$.
Hence, assuming there is no \ac{ISI}, $\xi=1$ minimizes~$P_\mathrm{e}$.
In this case, an error occurs only if $b[i]=1$ and $n_\RX[i]=0$.
Hence, \cref{eq:myPeDef} simplifies to $P_\mathrm{e}=1/2\times\Pr(n_\RX[i]=0;\; b[i]=1)$ assuming $\Pr(b[i]=1)=1/2$, i.e., assuming equal a priori probabilities for symbols $0$ and $1$.
For Poisson random variable $n_\RX[i]$, the average \ac{SER} simplifies to
\begin{equation}
    \label{eq:myPeApprox}
    P_\mathrm{e} = \frac{1}{2} \e^{-\overline{N}_\ob(t_0)},
\end{equation}
which we will refer to as the \emph{no \ac{ISI} approximation} in the following.

\section{Numerical Results}
\label{sec:numerical_results}
In this section, unless explicitly stated otherwise, we adopt the system parameters in \cref{tab:parameters}, where we use the viscosity of water for $\eta$, room temperature for $T_\f$, the saturation magnetization of magnetite for $\Msat$, the dimensions of a handheld magnet for $B_0$, $\Lmag$, and $\Rmag$, and typical values for microfluidic channels for $h$, $w$, and the central flow velocity $v_\f$~\cite{bruus_2007}.
Moreover, the \ac{MNP} parameters are chosen according to \cite{Zaloga_Development_2014}.
Thereby, for detection, we choose  $t_0=t_1$, where $t_1=d/v_\f$, i.e., the \ac{RX} takes a sample when particles are expected to arrive due to fluid flow.
\begin{table}
    \caption{%
        System Parameters.
    }
    \label{tab:parameters}
    \centering
    \begin{tabular}{llll}
        \toprule
        Parameter & Value & Parameter & Value \\
        \midrule
        $\eta$ & \SI{e-3}{\kilogram\per\meter\per\second} & $d$ & \SI{1}{\milli\meter} \\
        $T_\f$ & \SI{300}{\kelvin} & $h$ & \SI{10}{\micro\meter}\\
        $\ka$ & \SI{0.1}{\micro\meter\per\second} & $c_x$ & \SI{0.1}{\milli\meter}\\
        $\mh$ & \SI{27.5}{\nano\meter} & $c_z$ & \SI{1}{\micro\meter}\\
        $\sh$ & \SI{3}{\nano\meter} & $v_\f$ & \SI{0.5}{\milli\meter\per\second} \\
        $\Msat$ & \SI{4.75e5}{\ampere\per\meter} & $T$ & \SI{2}{\second} \\
        $B_0$ & \SI{1}{\tesla} & $\xi$ & $1$ \\
        $D$   & \SI{8e-12}{\meter^2\per\second} & $\vm$ & \SI{1}{\micro\meter\per\second} \\
        $\zeta$ & \SI{5e-10}{\kilogram\per\second} & $w$ & \SI{10}{\micro\meter} \\
        $\Lmag$ & \SI{5}{\centi\meter} & $c_y$ & \SI{1}{\micro\meter} \\
        $\Rmag$ & \SI{0.5}{\centi\meter} & $t_0$ & \SI{2}{\second} \\
        $\dm$ & \SI{5}{\milli\meter} & $\Ntx$ & \num{e3} \\
        $\Vs$ & \SI{4.5e-2}{\nano\meter^3} & $\Cs$ & \SI{1.23e-3}{\nano\meter^{-3}} \\
    \bottomrule
\end{tabular}
\end{table}
%
\changed[I:6]{We note that, in this section, we consider an offline performance evaluation where computational complexity is not an issue. Thereby, the computational complexity of our analysis increases linearly with $N$ because $\Pob(t)$ in \eqref{eq:received_in_tube} is the product of two summations over $N+1$ terms via $\Poby(t)$ in \eqref{eq:Poby} and $\Pobz(t)$ in \eqref{eq:constant_bounded_exact}, respectively.}

For simulation of the system described in \cref{sec:system_model}, we use a particle-based approach where time advances in discrete time steps $\Delta t$ and the position of each particle is tracked and updated in each time step, see e.g.~\cite[Eq.~(1)]{farsad_comprehensive_2016}.
\changed[II:2]{We use particle-based simulation since it captures the random arrivals of particles at the receiver which is crucial for performance analysis of \ac{MC} systems~\cite{noel_accord}.%
}
Then, for the received signal, for each time step, the number of particles within the receiver volume is counted.
Within each simulation step, if a particle crosses a channel boundary, it is removed from the simulation environment with probability $\Pad$ and reflected back into the channel with probability $1-\Pad$. 
\changed[I:1]{For reflection, in our simulation algorithm, we employ a perfect elastic reflection~\cite[Appendix~B.1]{noel_accord}, i.e., if an updated particle-position lies outside of the assumed boundaries then the new position is chosen as the respective mirror point.}
Qualitatively, $\Pad$ is related to the adsorption coefficient $\adcoeff$ via the prescribed flux at the boundary in \eqref{eq:boundary_flux}.
A more careful analysis reveals that the relationship between the adsorption coefficient $\adcoeff$ and the adsorption probability $\Pad$ in a particle-based simulation with simulation time step $\Delta t$ is not straightforward to obtain~\cite{andrews_accurate_2009}.
Nevertheless, this relation can be determined numerically.
For convenience, the following polynomial fit for flat boundaries was proposed in \cite[Eq.~(29)]{andrews_accurate_2009} and was also used in our simulations:
\begin{equation}
    \Pad = \madcoeff\sqrt{2\pi} - \num{3.33321} \madcoeff^2 + \num{3.35669} \madcoeff^3 - \num{1.52092} \madcoeff^4,
\end{equation}
where $\madcoeff=\Big[\adcoeff\sqrt{\frac{\Delta t}{2D}}\Big]_0^1$ and $[s]_a^b=\min\{\max\{s,a\},b\}$.

\begin{figure}[!t]
    \centering
    \subfloat[]{
        \includegraphics{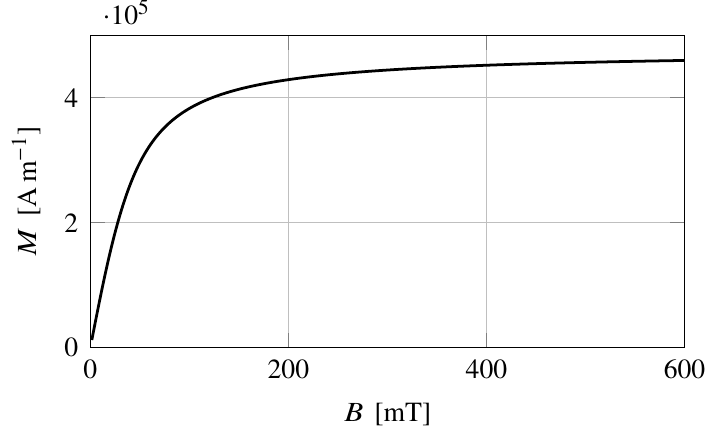}
        \label{fig:hysteresis}
    }
    \hfil
    \subfloat[]{
        \includegraphics{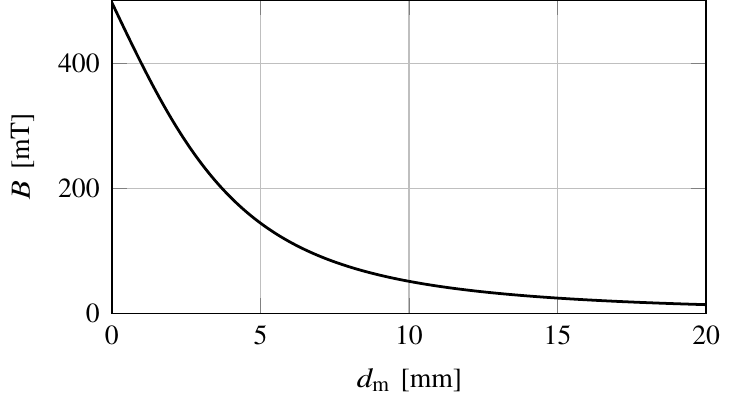}
        \label{fig:fluxDens}
    }
    \caption{
        (a) Magnetization curve of SPIONs according to \eqref{eq:hysteresis} and (b) magnetic field of a cylindrical magnet along its symmetry axis according to \eqref{eq:Bz}.
    }
    \label{fig:magnetism}
\end{figure}
In Fig.~\ref{fig:magnetism}, we evaluate the magnetic properties of the considered system.
In particular, in \cref{fig:hysteresis}, we evaluate the particle magnetization of a single SPION according to \cref{eq:hysteresis} as a function of the applied magnetic field $B$ and in \cref{fig:fluxDens}, we evaluate the magnetic flux density of a cylindrical magnet along its symmetry axis according to \cref{eq:Bz}.

As $M(B)$ in \cref{eq:hysteresis} is point symmetric, in \cref{fig:hysteresis}, we only show $M(B)$ for $B\geq 0$.
We see that $M(B)$ is an increasing function which saturates to $\Msat$ for sufficiently large $B$.
Here, magnetization begins to saturate for $B= \SI{100}{\milli\tesla}$.
On the other hand, the magnetic field as a function of the distance to the surface of the magnet is a decreasing function which approaches 0 for sufficiently large distances and is finite on the surface of the magnet.
Furthermore, we can observe that the magnetic field decays rapidly as the distance to the magnet increases and changes significantly on the length scale of a few millimeter.
We observe that $B(\dm=\SI{5}{\milli\meter}) > \SI{100}{\milli\tesla}$.
Therefore, for $\dm<\SI{5}{\milli\meter}$, we expect $M(B)$ of the SPIONs within an \ac{MNP} to be saturated.
Hence, for the considered system parameters, we expect the magnetic drift velocity $\vm$ to be well approximated by \eqref{eq:vmBlarge}. 

\begin{figure}[!t]
    \centering
    \includegraphics{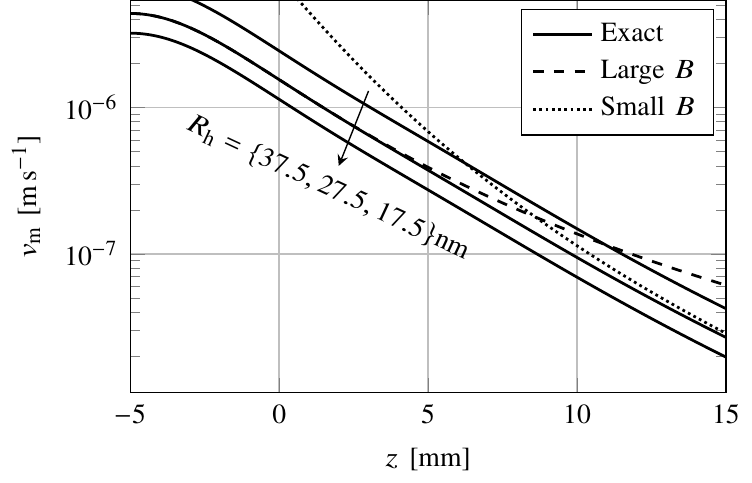}
    \caption{
        Induced magnetic drift velocity $\vm$ \eqref{eq:myF} as a function of the coordinate $z$ for $\dm=\SI{5}{\milli\meter}$.
    }
    \label{fig:driftVel}
    \vspace*{-5mm}
\end{figure}
In Fig.~\ref{fig:driftVel}, we show the induced drift velocity $\vm$ in \eqref{eq:myF} as a function of $z$ in and around the microfluidic channel for $\dm=\SI{5}{\milli\meter}$.
The drift velocity applies to \acp{MNP} placed within the magnetic field shown in Fig.~\ref{fig:fluxDens} where each embedded SPION is magnetized as shown in Fig.~\ref{fig:hysteresis}.
Eq.~\eqref{eq:myF} is evaluated for $\Rhyd\in\{\mh,\mh\pm\SI{10}{\nano\meter}\}$.
Furthermore, for $\Rhyd=\mh$, we also show approximations \eqref{eq:vmBlarge} and \eqref{eq:vmBsmall} generally valid for very large and very small magnetic fields, respectively.
The fast decay of $B$ over distance (see \cref{fig:fluxDens}) gives rise to a rapid decay of the induced magnetic drift velocity.
As the magnetic field is large close to the magnet and decays with distance, for short distances, approximation \eqref{eq:vmBlarge} assuming $M(B)=\Msat$ is most accurate.
Similarly, as the magnetic field decays to zero far away from the magnet, for larger distances, the approximation of the drift velocity in \eqref{eq:vmBsmall} matches the actual curve well.
In the intermediate regime (e.g., at position $z=\SI{7}{\milli\meter}$), neither approximation is very accurate.
Overall, the approximation for a large magnetic field is more reasonable for the considered distance of $\dm=\SI{5}{\milli\meter}$ (here at $z=\SI{0}{\milli\meter}$) and height $h=\SI{10}{\micro\meter}$ in Table~\ref{tab:parameters} where $\vm\approx \SI{1}{\micro\meter\per\second}$ is obtained which we will use in the following.
Here, the magnetic field gradient in \eqref{eq:dBz} can be evaluated as $-B'(h/2)=\SI{35.23}{\tesla\per\meter}$ at a magnetic flux density of $B=\SI{144}{\milli\tesla}$.
In general, for the considered setup, even for $z=\SI{-5}{\milli\meter}$ on the surface of the magnet, $\vm<\SI{10}{\micro\meter\per\second}$ which is much smaller than the considered flow velocity $\vf=\SI{0.5}{\milli\meter\per\second}$.
As is evident from \eqref{eq:myF}, a change in $\Rhyd$ simply scales the drift velocity.
Moreover, since $\vm$ changes with $z$ on the order of millimeter, the drift velocity can be reasonably considered as constant within the microfluidic channel of height $h=\SI{10}{\micro\meter}$, i.e., for $z\in[0,\SI{10}{\micro\meter}]$.

\begin{figure}[!t]
    \centering
    \subfloat[]{
        \includegraphics{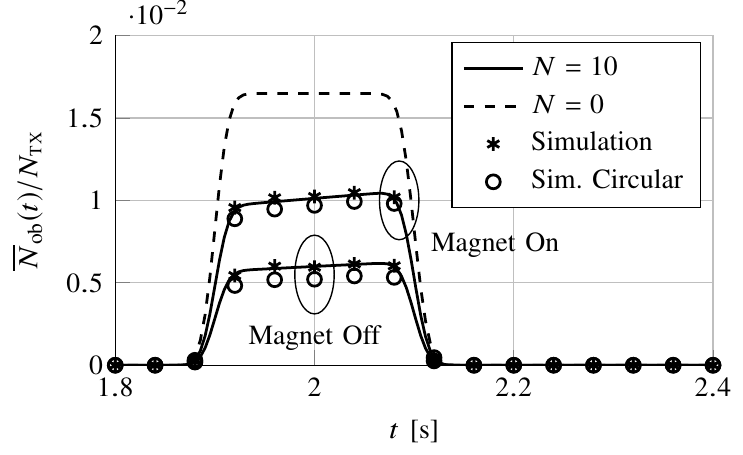}
        \label{fig:ir_reflect}
    }
    \hfil
    \subfloat[]{
        \includegraphics{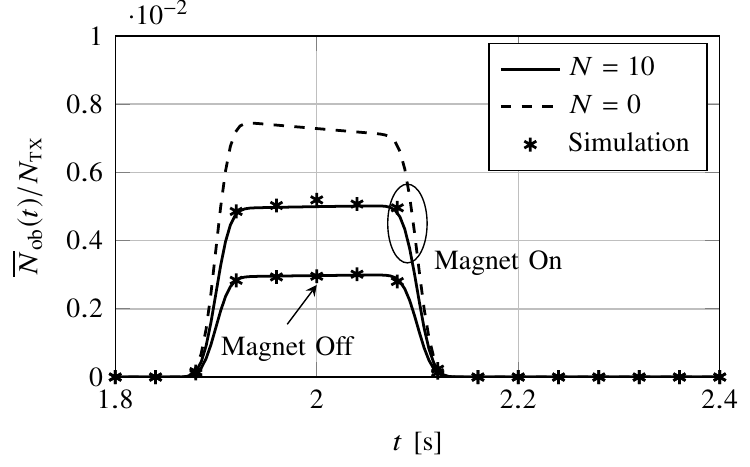}
        \label{fig:ir_absorb}
    }
    \caption{
        Impulse response $\Nobmean(t)/\Ntx$ in \cref{eq:myNob} for the magnet being turned on and off.
        (a) Impulse response for $\ka=0$.
        (b) Impulse response for $\ka=\SI{0.1}{\micro\meter\per\second}$.
        Simulation results with $\Delta t=\SI{2}{\milli\second}$ have been averaged over $10^3$ realizations.
        \changed[I:4]{The flow velocity is set to $\vf=\SI{0.5}{\milli\meter/\second}$.}
        \vspace*{-5mm}
    }
    \label{fig:ir}
\end{figure}
In \cref{fig:ir}, we show $\Nobmean(t)/\Ntx$ for $\Ntx=1000$ averaged over \num{e3} independent realizations for times around $t_1=\SI{2}{\second}$.
Analytical results are shown for nominal particle sizes, whereas the simulations were performed for log-normal distributed particle sizes.
Thereby, in \cref{fig:ir_reflect} and \cref{fig:ir_absorb}, fully reflective boundaries ($\ka=0$) and partially adsorbing boundaries ($\ka>0$) are considered, respectively.
Results are shown for the magnet being \changed{turned} on ($\vm=\SI{1}{\micro\meter\per\second}$) and the magnet being turned off ($\vm=0$).
\changed[II:5]{The latter case is equivalent to a conventional \ac{MC} system employing non-magnetic particles.}
For clarity, the long-time approximation (i.e., $N=0$) is shown only for the magnet being turned on.
Moreover, for comparison, in \cref{fig:ir_reflect}, simulation results for a circular duct, which might serve as a simple model for blood vessels, are also shown where the radius was chosen as $\sqrt{wh/\pi}$, i.e., the area of the cross section is the same as for the rectangular duct.
Results for a circular boundary are only shown for fully reflective boundaries where simulation%
\footnote{%
    We study a circular boundary only by simulation because mathematical analysis of diffusion within a cylindrical shape with a vertical force component is very involved, and the obtained analytical results are less insightful than in the rectangular case \cite{Schaefer_Transfer_2018}.
}
is straightforward since the adsorption probability is 0.

\cref{fig:ir} shows that turning the magnet on increases the number of observed \acp{MNP}.
There is a time window of approximately $\SI{0.2}{\second}$ centered around $t_1=\SI{2}{\second}$ within which a nonzero number of \acp{MNP} can be observed independent of the magnetic field gradient.
Hence, for a symbol interval size of $T=\SI{2}{\second}$, \ac{ISI} does not play a significant role for the given parameters due to the flow-dominated transport of particles%
\footnote{\label{fn:isi}%
    \changed[I:5]{We note that, in general, ISI is an important issue in \ac{MC}~\cite{farsad_comprehensive_2016}. However, though not excluded in our analysis, for the considered system parameters, ISI is negligible. We note also that ISI will be more severe in case of laminar flow, the study of which constitutes an interesting topic for future work~\cite{wicke_duct_2017}.}%
}.

The equilibrium approximation ($N=0$) for $\vm=\SI{1}{\micro\meter\per\second}$, overestimates the number of observed particles as within the time frame where \acp{MNP} can be observed, the steady state has not been reached yet.
Overall, \cref{fig:ir} confirms that magnetically targeting the \ac{RX} is effective in increasing the number of observed \acp{MNP}.
The simulation for a circular tube of the same cross sectional area yields similar results as the rectangular case but overall exhibits a slightly smaller number of observed \acp{MNP}.
This can be explained by the particles having less space to settle close to the \ac{RX} due to the curvature and hence being reflected more often at the boundary.
Overall, the width of the impulse response depends on the \ac{RX} size and the flow velocity $\vf$ which cannot be controlled.
On the other hand, the amplitude of the impulse response can be influenced by the magnetic field.
In the considered case, the particle size distribution does not have a large effect on the impulse response as the simulation results taking into account this distribution match the analytical curves, which consider only the nominal particle size.

\begin{figure}[!t]
    \centering
    \includegraphics{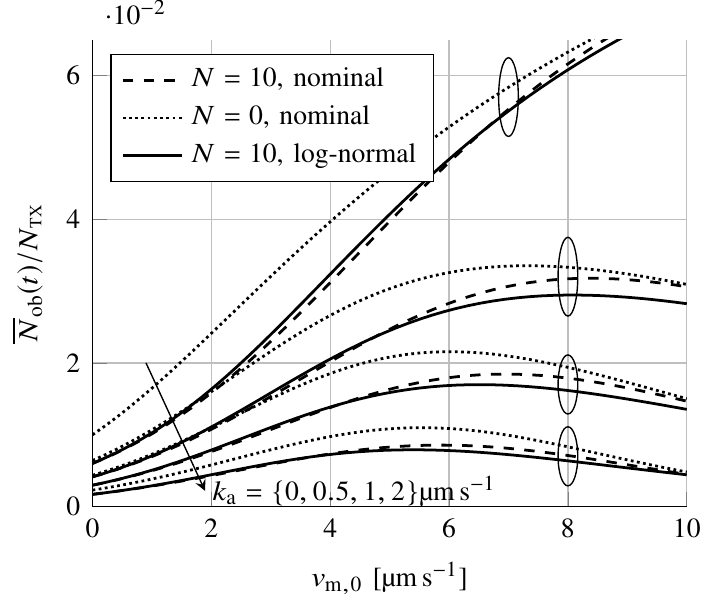}
    \caption{
        Impulse response $\Nobmean(t_0)/\Ntx$ in \eqref{eq:myNob} as a function of the magnetic drift velocity $\vmnom$ for nominal and log-normal distributed particle sizes and different numbers of terms in the solutions \eqref{eq:constant_bounded_exact} and \eqref{eq:Poby}.
        The initial release point is set to $z_0=h$.
        \changed[I:4]{The flow velocity is set to $\vf=\SI{0.5}{\milli\meter/\second}$.}
        \vspace*{-5mm}
    }
    \label{fig:signalStrength}
\end{figure}
In Fig.~\ref{fig:signalStrength}, we further investigate the achievable amplitude of the impulse response when varying the magnetic drift velocity $\vmnom$ from \SIrange{0}{10}{\micro\meter\per\second}.
In particular, we show the value of the impulse response $\Nobmean(t_0)/\Ntx$ at the sampling point $t=t_0$ as a function of the nominal magnetic drift velocity $\vmnom$ for different adsorption coefficients and for $N=0$ as well as $N=10$ in \eqref{eq:myNob}.
Thereby, both nominal and log-normal distributed particle sizes are taken into account for $\vm$ and $D$ via \eqref{eq:vmlog} and \eqref{eq:Dlog}.
In total, $\Ntx=\num{e6}$ realizations of $\vm$ and $D$ in $P_{\ob,i}(t)$ in \eqref{eq:myNob} are considered.
Furthermore, the approximation for $N=0$ in \eqref{eq:Pob_eq_approx} is shown which captures the asymptotic behavior.
In general, for all curves there exists a maximum which can be explained as follows.
For very large $\vmnom$, particles get immediately dragged towards the lower boundary and there, over time, interact more often with the boundary and thus have a higher chance of adsorption leading to a smaller received signal.
On the other hand, for very small $\vm$, particles stay for a prolonged time close to the initial position $z_0=h$ where they have been released and possibly do not reach the \ac{RX} close to position $z=0$.
Hence, an intermediate $\vmnom$ appears optimal.
It can also be observed that the maximum increases and moves to larger $\vmnom$ for decreasing adsorption coefficients.
This behavior is expected as for smaller $\adcoeff$, more particles can be dragged towards $z=0$ without a loss by adsorption.
Moreover, the deviation between the approximation for large times ($N=0$) and the curve for $N=10$ decreases for larger $\vmnom$ and for larger $\adcoeff$ since in both cases the quasi steady-state is reached earlier because the impact of magnetic drift and adsorption increases compared to that of diffusion.
Overall, the equilibrium approximation describes the behavior qualitatively well.
Furthermore, the existence of a maximum hints that in an application $\vmnom$ should be optimized depending on the adsorption coefficient.

Due to the log-normal particle size distribution there is some deviation from the nominal impulse response as $v_\m$ depends on the particle size.  
In particular, when $\vmnom$ is relatively small (e.g. $\vmnom=\SI{2}{\micro\meter\per\second}$) and large  (e.g. $\vmnom=\SI{8}{\micro\meter\per\second}$), more and fewer \acp{MNP} are observed at the \ac{RX} than expected based on the nominal impulse response, respectively. 
This can be explained as follows.
The nominal impulse response is obtained based on the assumption that the radius $\Rhyd$ of all \acp{MNP} is equal to the mean radius $\mh$.
In reality, the \ac{MNP} sizes are log-normal distributed, i.e., for some particles $\Rhyd<\mh$ and for others $\Rhyd>\mh$.
Thereby, $\vm$ increases with $\Rhyd$.
Hence, for small $\vmnom$, both particles with $\Rhyd=\mh$ and particles with $\Rhyd<\mh$ experience a small $\vm$, i.e., the probability of observing particles with $\Rhyd\leq\mh$ is uniformly small.
However, for small $\vmnom$, particles with $\Rhyd>\mh$ experience a larger $\vm$ and hence, contribute to an increased number of observed \acp{MNP}.
On the other hand, for large $\vmnom$, the magnetic force experienced by \acp{MNP} having radius $\Rhyd\geq\mh$ is uniformly relatively strong and almost all \acp{MNP} with $\Rhyd=\mh$ arrive at the \ac{RX}.
However, having smaller \acp{MNP} with $\Rhyd<\mh$, which experience a weaker magnetic force, decreases the number of observed \acp{MNP}.
In summary, for the impulse response there is a trade-off in amplitude between dragging the particles towards the boundary-mounted receiver and the increased loss of particles due to adsorption at the boundaries.

\begin{figure}
    \centering
    \includegraphics{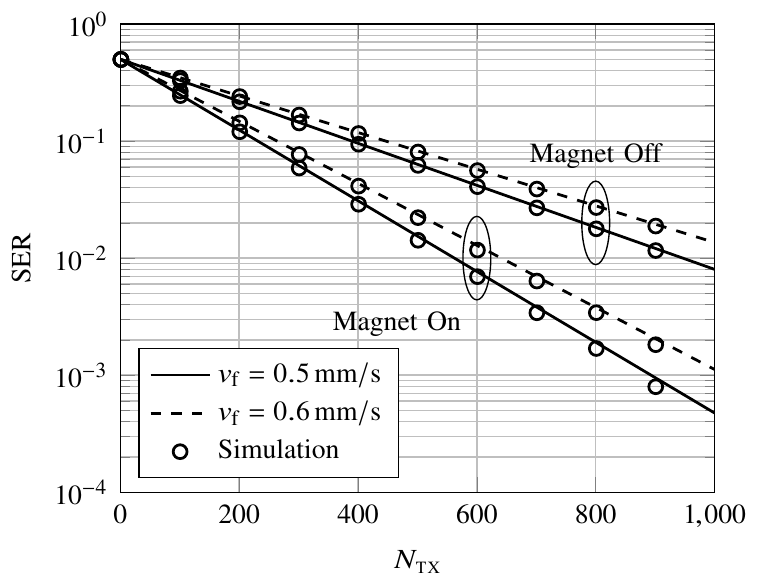}
    \caption{\label{fig:ser}
        Symbol error rate \changed{(SER)} as a function of the available number of \acp{MNP} per symbol.
        $P_\mathrm{e}$ in \cref{eq:myPeDef} with $K=10$ is shown for two different fluid flow velocities and with the magnet turned on and off. 
        Simulation results with $\Delta t=\SI{1}{\milli\second}$ have been averaged over $10^4$ independent realizations.
        Log-normal distributed particle sizes are considered.
        The initial release point is set to $z_0=h$.
    }
\end{figure}

In \cref{fig:ser}, we evaluate the symbol error rate when the magnet is turned on and off, respectively, as a function of the number of \acp{MNP} used per transmit pulse.
In particular, for each $N_\TX$, we show the \ac{SER} according to \cref{eq:myPeApprox} with $\Nobmean(t)$ from \eqref{eq:myNob}.
Simulation results are provided to verify the applied approximations.
Two different flow velocities, $\vf=\SI{0.5}{\milli\meter\per\second}$ and $\vf=\SI{0.6}{\milli\meter\per\second}$, are considered to evaluate the effect of possible flow variations.
Thereby, we have $t_0=d/\vf$ while $T=\SI{2}{\second}$ is fixed.
As for $T=\SI{2}{\second}$ no \ac{ISI} is expected for the chosen system parameters, cf.~\cref{fig:ir}, the no \ac{ISI} approximation in \cref{eq:myPeApprox} matches \cref{eq:myPeDef} which for clarity is not shown.
Furthermore, when the magnet is turned on, the \ac{SER} decreases more rapidly with $\Ntx$ since $\Nobmean(t_0)$ in \eqref{eq:myPeApprox} is larger compared to when the magnet is turned off.
We can also observe that the system is very sensitive to changes in the fluid flow which cannot be controlled externally.
Thereby, for
\changed[I:2]{$\vf=\SI{0.6}{\milli\meter/\second}$ the \ac{SER} is higher than for $\vf=\SI{0.5}{\milli\meter/\second}$ as, in this case, the time period $d/\vf$ in which the particles reach the center $x$-coordinate of the \ac{RX} is shorter and hence, before detection}
particles have less time to settle at $z=0$ where the \ac{RX} is mounted%
\footnote{\label{fn:ser_isi}%
    \changed[I:3,I:5]{We note that for a flow velocity of $\vf\ll\SI{0.5}{\milli\meter/\second}$, the \ac{SER} will increase due to a decreased received signal strength caused by diffusion along the $x$-axis. Furthermore, in this case, unless the symbol duration $T$ was significantly increased, we would also expect significant \ac{ISI}, i.e., \eqref{eq:myPeDef} would have to be used for evaluating the \ac{SER} instead of the no \ac{ISI} approximation in \eqref{eq:myPeApprox}.}%
}.
\changed[I:2]{This results in a decreased signal strength and hence in an increased \ac{SER}.}
However, overall, we note that turning the magnet on reduces the \ac{SER} significantly.
Therefore, employing \acp{MNP} for communication inside a microfluidic channel and controlling them by a magnet can significantly improve reliability despite a disruptive fluid flow.
This \ac{MNP} guidance may only be necessary at the receiver site and thus does not entail a large overhead in terms of additional required hardware.
Furthermore, the applied Poisson approximation is accurate for all considered $\Ntx$.

\begin{figure}[ht]
    \centering
    \includegraphics{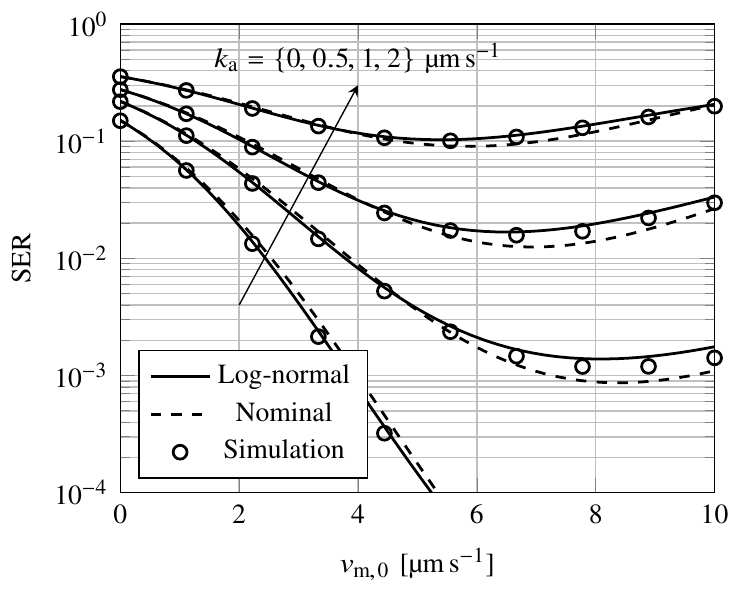}
    \caption{
        Symbol error rate (SER) as a function of the nominal magnetically induced drift velocity $\vmnom$.
        $P_\mathrm{e}$ in \cref{eq:myPeApprox} with $K=10$ is shown for different adsorption coefficients $\adcoeff$. 
        Simulation results with $\Delta t=\SI{1}{\milli\second}$ have been averaged over $10^4$ independent realizations.
        Nominal and log-normal distributed particle sizes are considered.
        The initial release point is set to $z_0=h$.
        \changed[I:4]{The flow velocity is set to $\vf=\SI{0.5}{\milli\meter/\second}$.}
        \vspace*{-5mm}
    }
    \label{fig:ser_vel}
\end{figure}

In Fig.~\ref{fig:ser_vel}, we plot the \ac{SER} as a function of the nominal magnetically induced drift velocity for $\Ntx=\num{e3}$ and for different adsorption coefficients $\adcoeff=\{0,0.5,1,2\}\,\si{\micro\meter/\second}$.
Both nominal and log-normal distributed particle sizes are considered.
Furthermore, simulation results are provided to validate the analysis.
We can observe that the \ac{SER} depends strongly on the adsorption coefficient.
For small adsorption coefficients, e.g., for $\adcoeff=\SI{0}{\micro\meter/\second}$, an increase in the magnetically induced drift velocity $\vm$ decreases the \ac{SER}.
For larger adsorption coefficients, e.g., for $\adcoeff=\SI{1}{\micro\meter/\second}$, there exists an optimal $\vmnom$ which minimizes the \ac{SER} due to the trade-off between dragging particles close to the \ac{RX} and losing particles due to adsorption.
Thereby, deviations due to the particle size distribution are most severe for larger drift velocities, e.g., at $\vmnom=\SI{8}{\micro\meter\per\second}$.
In general, the behavior of the curves can be understood as a scaled version of the curves shown in Fig.~\ref{fig:signalStrength} due to the logarithmic scale and the exponential behavior of the \ac{SER} in \eqref{eq:myPeApprox}.
Hence, for employing \acp{MNP} successfully $\vmnom$ needs to be optimized, i.e., for a given adsorption coefficient, the \ac{SER} can be used as design criterion for the magnetic field.

\section{Conclusion}
\label{sec:conclusion}
In this paper, we proposed the use of \acp{MNP} as information carriers for \ac{MC} systems.
In particular, we showed how the movement of \acp{MNP} can be modeled as diffusion with drift.
To this end, we studied the magnetic drift velocity resulting from a magnetic force caused by a magnetic field gradient.
Thereby, we highlighted the dependence of the drift velocity and the diffusion coefficient on the particle size.
Subsequently, we introduced a technique to solve the diffusion equation with drift in a bounded environment and applied this technique to derive the impulse response of a microfluidic channel subject to fluid flow, diffusion, and magnetic drift.
Moreover, we showed how the particle size distribution can be incorporated in the impulse response.
By numerical evaluation, we illustrated how a log-normal particle size distribution and boundary adsorption affect the impulse response for different magnetic field gradients.
Thereby, we found a trade-off between attracting particles towards the \ac{RX} and adsorption at the boundaries.
Finally, by evaluating the \ac{SER}, we investigated the sensitivity of the system to variations in the fluid flow velocity and found that applying an external magnetic field can ensure reliable communication.
Hence, the use of \acp{MNP} as information carriers is attractive for application in \ac{MC} systems operating in artificial or natural microfluidic environments such as blood vessels.
\changed[II:6]{The theoretical analysis presented in this paper can provide guidelines for designing such systems.}

\appendix[Derivation of PDF $p_z(z;t)$]
Solutions to the one-dimensional diffusion equation without drift, which is mathematically referred to as the \emph{heat equation}, are well known for various boundary conditions~\cite{carslaw_conduction_1986}.
Motivated by this, using a variable substitution and separation of variables in \eqref{eq:pde_bounded}, we obtain an equivalent problem formulation in terms of an auxiliary function $q(z;t)$ without drift term, i.e., the heat equation, but with modified boundary conditions~\cite{perez_guerrero_analytical_2009}.
To this end, we implicitly define $q(z;t)$ as
\begin{equation}
    \label{eq:p_transform}
    p_z(z;t) = q(z;t)\exp\left(-u(z - z_0) - 
    Du^2t\right),
\end{equation}
where $u=\vm/(2D)$.
Substituting \cref{eq:p_transform} in \eqref{eq:pde_bounded}, for $0<z<h$ and $t>0$, we obtain the following \ac{PDE} with boundary and initial conditions in terms of $q(z;t)$
\begin{IEEEeqnarray}{rCl+l}
\label{eq:auxiliary_bounded}
\D{t} q(z;t) &= &D\D[2]{z} q(z;t), & \IEEEyesnumber\IEEEyessubnumber*\\
\frac{\partial}{\partial z} q(z;t)     &= &(\mcoeff - u)q(z;t), &z=0 \\
\frac{\partial}{\partial z} q(z;t)     &= &-(\mcoeff + u)q(z;t), &z=h \label{eq:aux_bound_h}\\
q(z;t) &= &\delta(z-z_0), &t = 0.\label{eq:auxiliary_bounded_initial}
\end{IEEEeqnarray}

Function $q(z;t)$ in \eqref{eq:auxiliary_bounded} is separable in $z$ and $t$ and therefore the auxiliary function $q(z;t)$ can be expressed as a series \cite{carslaw_conduction_1986}
\begin{equation}
    \label{eq:qSeries}
    q(z;t) = \sum_{n=-\infty}^\infty Z_n(z) \exp(-D s_n^2 t) a_n,  
\end{equation}
with coefficients $a_n$.
Thereby, $s_n$ and $Z_n(z)$ in \eqref{eq:qSeries} need to satisfy the \emph{eigenproblem}
\begin{IEEEeqnarray}{rCl+l}
\label{eq:eigenproblem}
    \D[2]{z} Z_n(z)  &= &- s_n^2 Z_n(z), & 0<z<h \IEEEyesnumber\IEEEyessubnumber* \\
    \D{z} Z_n(z) &= &(\mcoeff - u)Z_n(z), & z = 0 \\
    \D{z} Z_n(z) &= &-(\mcoeff + u)Z_n(z), & z = h.
\end{IEEEeqnarray}
For a comprehensive solution, we have to consider all cases of $s_n^2>0$, and $s_n^2\leq0$, which correspond to $s_n$ being real and imaginary, respectively.

A particular solution of \eqref{eq:eigenproblem} is given by
\begin{equation}
    \label{eq:app_cossol}
    Z_n(z) = \cos(s_n z) + \frac{\mcoeff-u}{s_n}\sin(s_n z),
\end{equation}
where $s_n$ has to satisfy
\begin{equation}
    \label{eq:app_taneq}
    \tan(s_nh) = \frac{2s_n\mcoeff}{s_n^2 - (\mcoeff^2 - u^2)}.
\end{equation}
In fact, \eqref{eq:app_taneq} has infinitely many real solutions for any combination of $\mcoeff$ and $u$.
On the other hand, \eqref{eq:app_taneq} can have at most one imaginary solution $s_0=\im\sigma$ as will be seen in the following.
In this case, \eqref{eq:app_cossol} becomes
\begin{equation}
    \label{eq:app_coshsol}
    Z_0(z) = \cosh(\sigma z) + \frac{\mcoeff - u}{\sigma} \sinh(\sigma z),
\end{equation}
where by \eqref{eq:app_taneq} $\sigma>0$ is the solution of
\begin{equation}
    \label{eq:app_tanheq}
    \tanh(\sigma h) = \frac{-2\sigma \mcoeff}{\sigma^2 + (\mcoeff^2 - u^2)}.
\end{equation}
Eq.~\eqref{eq:app_tanheq} has exactly one solution if $u > u_\crit$ and none otherwise, where
\begin{equation}
    u_\crit = \sqrt{\frac{2}{h}\mcoeff + \mcoeff^2}.
\end{equation}
Finally, for $s_0\to0$, \eqref{eq:app_cossol} becomes
\begin{equation}
    Z_0(z) = 1+(\mcoeff-u)z
\end{equation}
if $u=u_\crit$.

Interestingly, it can be shown that \eqref{eq:app_taneq} has a solution $s_0\in(0,\pi/h)$ when $u<\ucrit$.
Hence, for $n=0$, we can distinguish between a trigonometric, a hyperbolic, and an affine function $Z_0(z)$ for $u<\ucrit$, $u>\ucrit$, and $u=\ucrit$, respectively.

Imposing the initial condition in \eqref{eq:auxiliary_bounded_initial} and exploiting the orthogonality of the proposed $Z_n(z)$, we find the coefficients $a_n\neq 0$ in \eqref{eq:qSeries} as
\begin{IEEEeqnarray}{rCl}
    a_n &=& \frac{Z_n(z_0)}{\norm{Z_n}^2}, \\
    \norm{Z_n}^2 &=& \int_0^h |Z_n(z)|^2\;\mathrm dz.
\end{IEEEeqnarray}
This completes the proof.

\bibliographystyle{IEEEtran}
\bibliography{main.bbl} 

\begin{IEEEbiography}
    [{\includegraphics[width=1in,height=1.25in,clip,keepaspectratio]{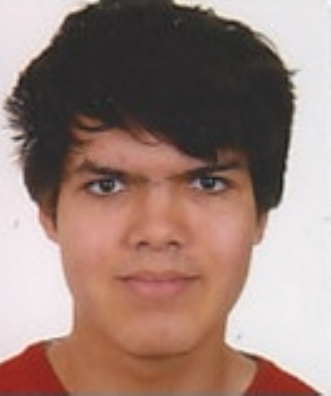}}]
    {Wayan Wicke}
    (S'17) was born in Nuremberg, Germany, in 1991.
    He received the B.Sc.\ and  M.Sc.\ degrees in electrical engineering from the Friedrich-Alexander University Erlangen-Nürnberg (FAU), Erlangen, Germany, in 2014 and 2017, respectively, where he is currently pursuing the Ph.D.\ degree.
    His research interests include  statistical signal processing and digital communications with a focus on molecular communication.
\end{IEEEbiography}
\begin{IEEEbiography}
    [{\includegraphics[width=1in,height=1.25in,clip,keepaspectratio]{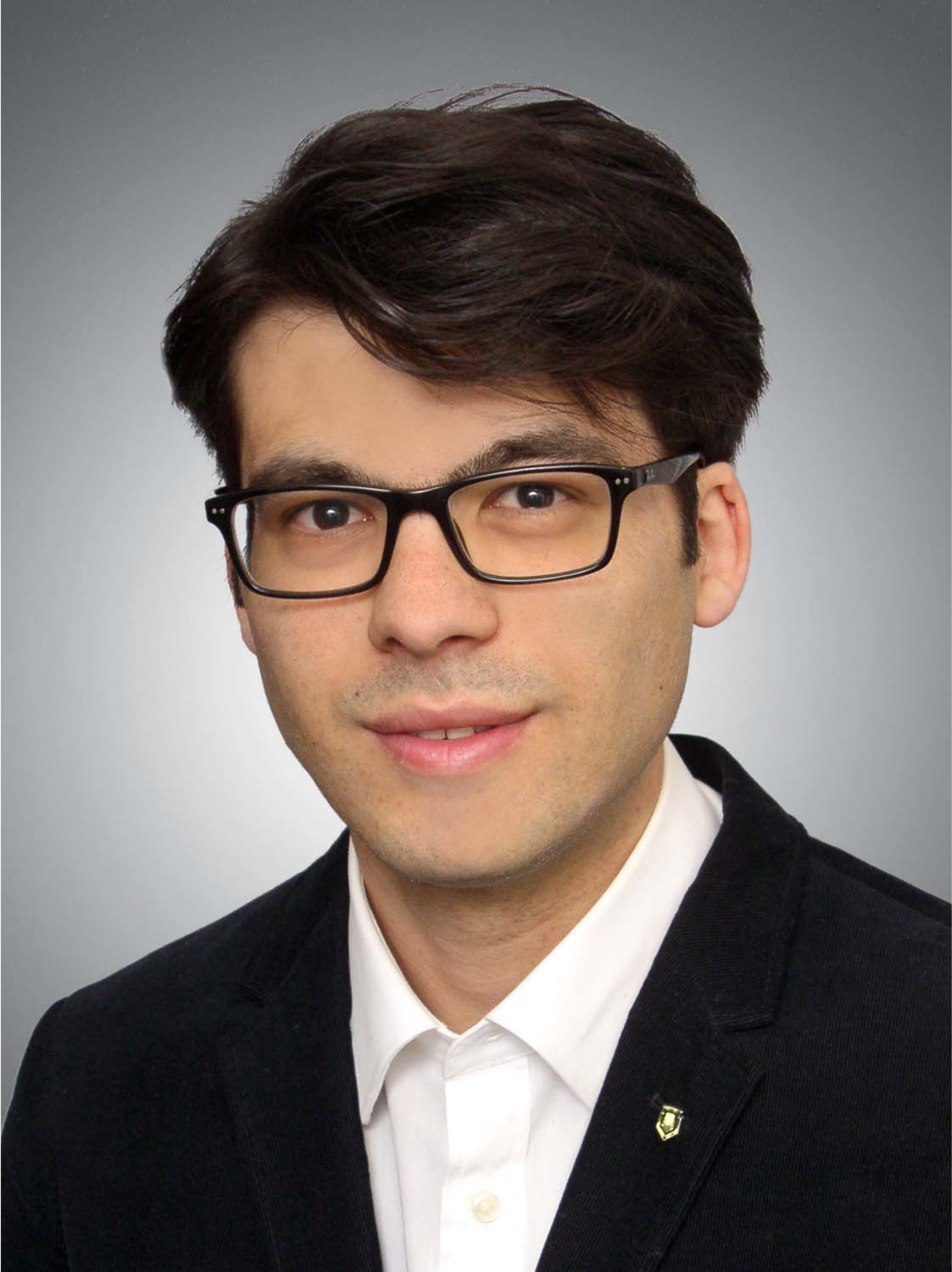}}]
    {Arman Ahmadzadeh}
    (S'14) received the B.Sc.\ degree in electrical engineering from the Ferdowsi University of Mashhad, Mashhad, Iran, in 2010, and the M.Sc.\ degree in communications and multimedia engineering from the Friedrich-Alexander University Erlangen-Nürnberg, Erlangen, Germany, in 2013, where he is currently pursuing the Ph.D.\ degree in electrical engineering with the Institute for Digital Communications.
    His research interests include physical layer molecular communications.
    Arman served as a member of Technical Program Committees of the Communication Theory Symposium for the IEEE International Conference on Communications (ICC) 2017 and 2018.
    Arman received several awards including the ``Best Paper Award'' from the IEEE ICC in 2016, ``Student Travel Grants'' for attending the Global Communications Conference (GLOBECOM) in 2017, and was recognized as an Exemplary Reviewer of the \textsc{IEEE Communications Letters} in 2016.
\end{IEEEbiography}
\begin{IEEEbiography}
    [{\includegraphics[width=1in,height=1.25in,clip,keepaspectratio]{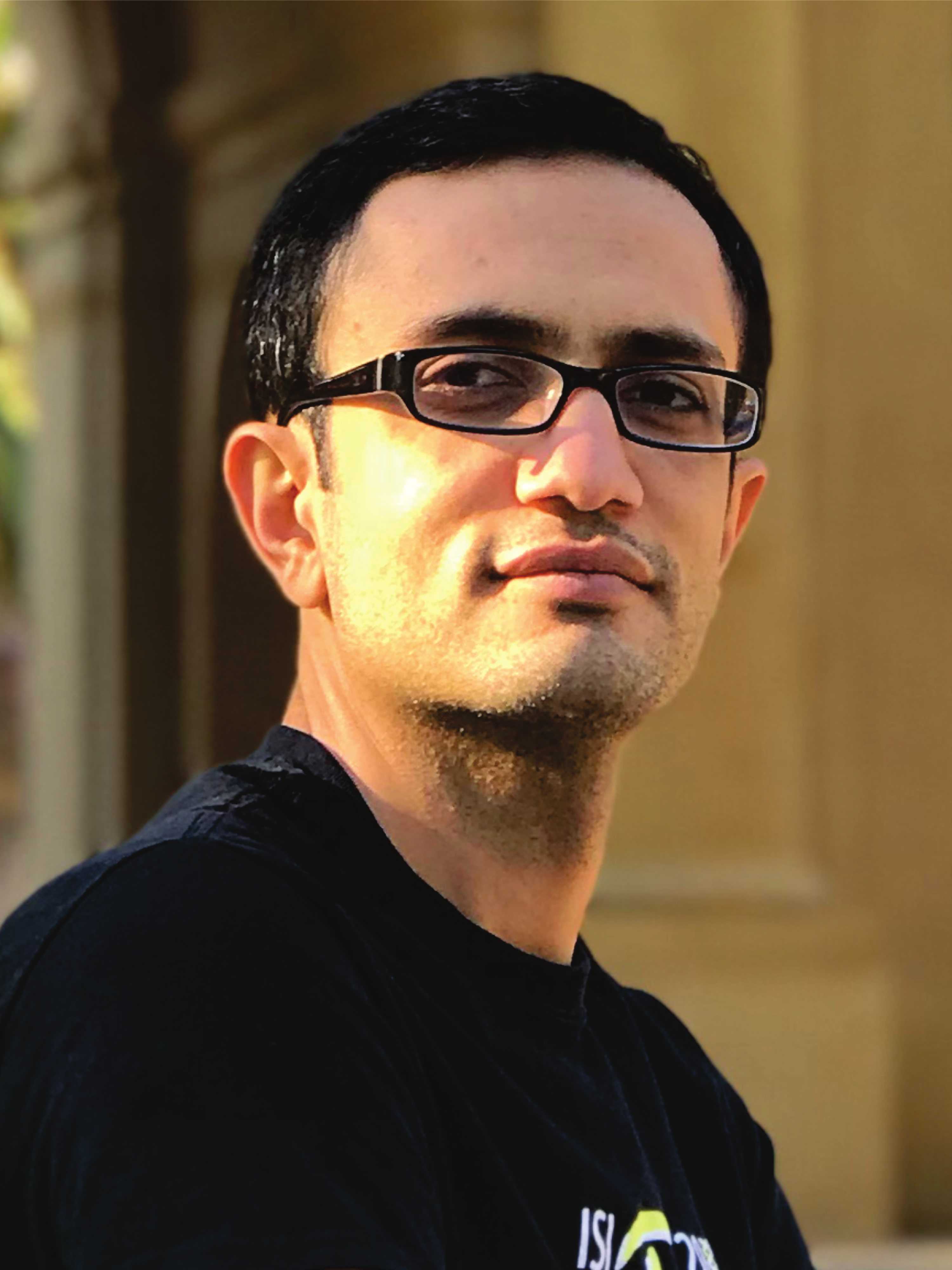}}]
    {Vahid Jamali}
    (S'12) received the B.S.\ and M.S.\ degrees (Hons.) in electrical engineering from the K.\ N.\ Toosi University of Technology, Iran, in 2010 and 2012, respectively.
    He is working toward his Ph.D.\ degree at the Friedrich-Alexander University Erlangen-Nürnberg (FAU), Erlangen, Germany.
    He was a visiting research scholar at the Stanford University, USA, in 2017.
    His research interests include wireless communications, molecular communications, multiuser information theory, and signal processing.
    He served as a member of Technical Program Committees and as a technical reviewer for several IEEE conferences and journals.
    He received several awards for his work including the Best Paper Award from the IEEE International Conference on Communications (ICC) in 2016, visiting scholarship from the German Academic Exchange Service (DAAD) in 2017, winner of the Best 3 Minutes (Ph.D.) Thesis (3MT) Presentation from the IEEE Wireless Communications and Networking Conference 2018, Goldener Igel Publication Award from FAU in 2018, Exemplary Reviewer Awards of the \textsc{IEEE Communications Letters} in 2014 and \textsc{IEEE Transactions on Communications} in 2017, and student travel grants for the SP Coding and Information School, Sao Paulo, Brazil in 2015, the Training School on Optical Wireless Communications, Istanbul, Turkey, in 2015, and the IEEE ICC 2017.
\end{IEEEbiography}
\begin{IEEEbiography}
    [{\includegraphics[width=1in,height=1.25in,clip,keepaspectratio]{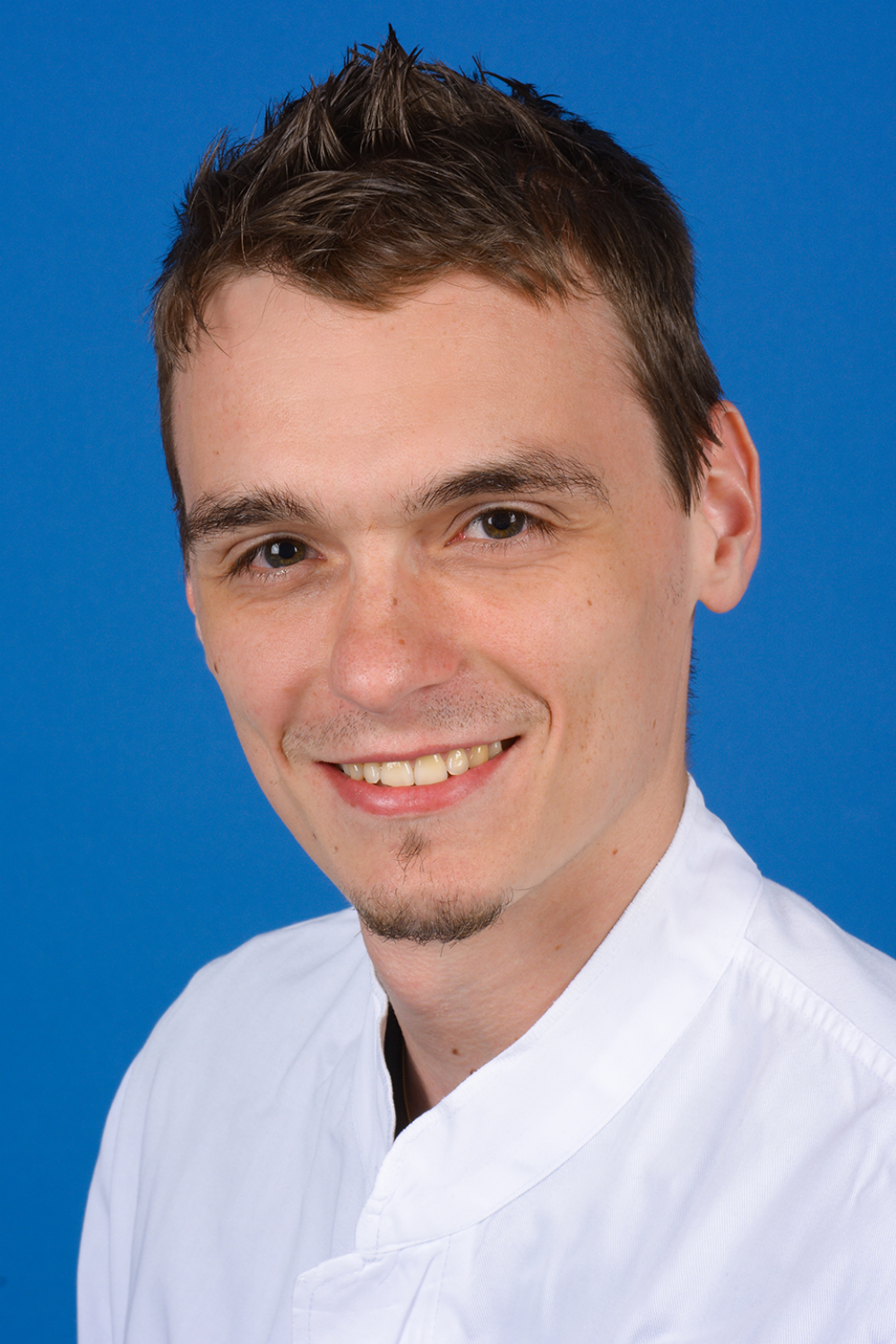}}]
    {Harald Unterweger}
    is a PostDoc and deputy head of the synthesis and analytics department of the Section of Experimental Oncology and Nanomedicine (SEON).
    His work focuses in the development and characterization of magnetic nanoparticles for biomedical and technical applications.
    Harald earned a M.Sc.\ degree in nanotechnology and a Ph.D.\ degree in material sciences from the Friedrich-Alexander University Erlangen-Nürnberg (FAU), Erlangen, Germany.
    For his Ph.D.\ thesis, he received the dissertation award from the German Ferrofluid Society and the dissertation award from the FAU's Technical Faculty (Freundeskreis der Alumni Technische Fakult\"at Erlangen).
\end{IEEEbiography}
\begin{IEEEbiography}
    [{\includegraphics[width=1in,height=1.25in,clip,keepaspectratio]{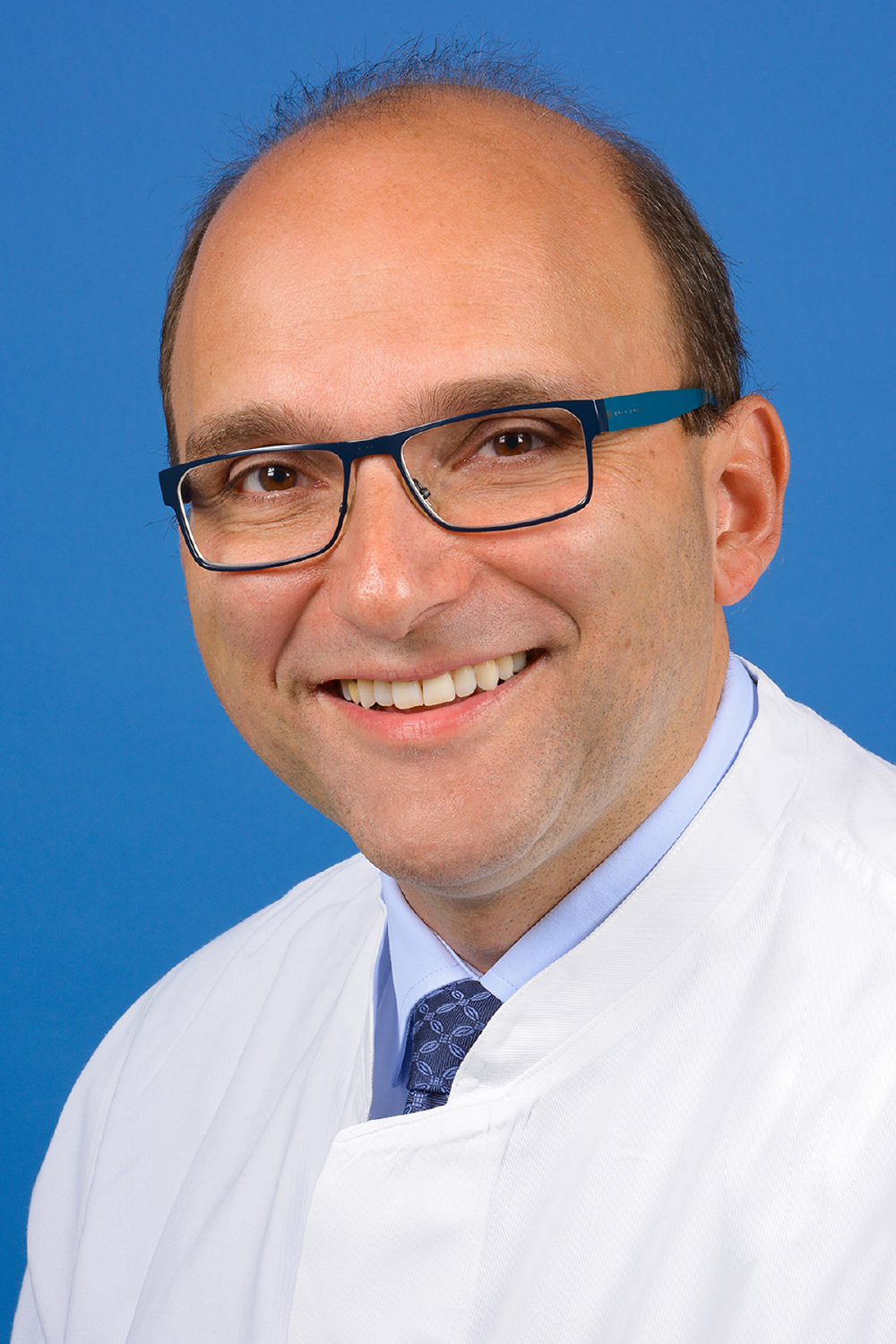}}]
    {Christoph Alexiou}
    received his Ph.D.\ in 1995 from the TU-Munich, Medical school and 2002 he changed to the ENT-Department in Erlangen, Germany, where he performed his postdoctoral lecture qualification (Habilitation).
    He is working there as an assistant medical director in the clinic and leads the Section for Experimental Oncology and Nanomedicine (SEON).
    Since 2009 he owns the W3-Else Kröner-Fresenius-Foundation-Professorship for Nanomedicine at the University Hospital Erlangen.
    His research is addressing the emerging fields of Diagnosis, Treatment, Regenerative Medicine and Molecular Communication using magnetic nanoparticles.
    He received for his research several national and international renowned awards.
\end{IEEEbiography}
\begin{IEEEbiography}
    [{\includegraphics[width=1in,height=1.25in,clip,keepaspectratio]{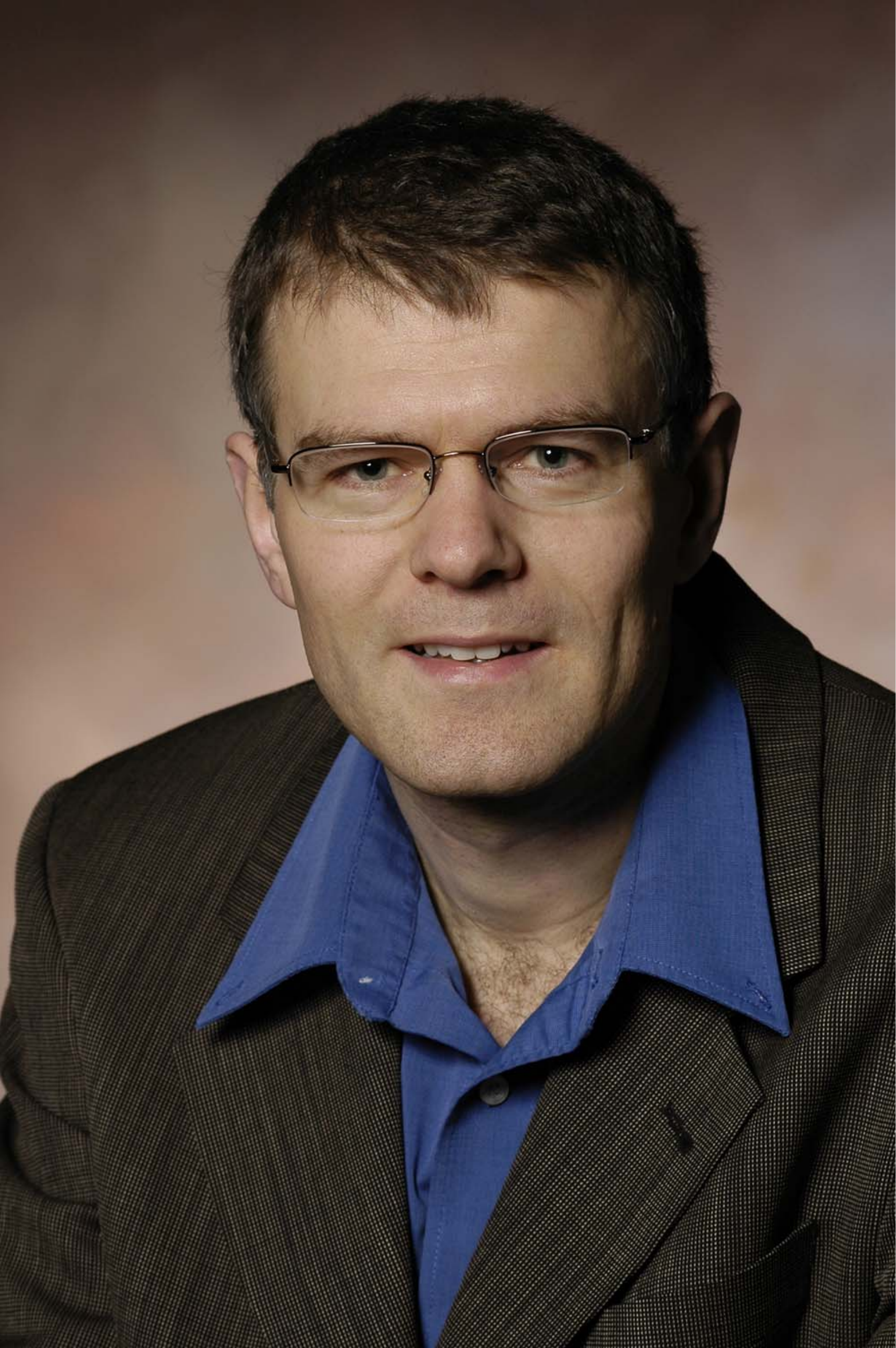}}]
    {Robert Schober}
    (S'98, M'01, SM'08, F'10) received the Diplom (Univ.) and the Ph.D.\ degrees in electrical engineering from the Friedrich-Alexander University Erlangen-Nürnberg, Erlangen, Germany, in 1997 and 2000, respectively.
    From 2002 to 2011, he was a Professor and Canada Research Chair at the University of British Columbia (UBC), Vancouver, Canada.

    Since January 2012 he is an Alexander von Humboldt Professor and the Chair for Digital Communication at FAU.
    His research interests fall into the broad areas of Communication Theory, Wireless Communications, and Statistical Signal Processing.

    Robert received several awards for his work including the 2002 Heinz Maier-Leibnitz Award of the German Science Foundation (DFG), the 2004 Innovations Award of the Vodafone Foundation for Research in Mobile Communications, a 2006 UBC Killam Research Prize, a 2007 Wilhelm Friedrich Bessel Research Award of the Alexander von Humboldt Foundation, the 2008 Charles McDowell Award for Excellence in Research from UBC, a 2011 Alexander von Humboldt Professorship, a 2012 NSERC E.W.R.\ Stacie Fellowship, and a 2017 Wireless Communications Recognition Award by the IEEE Wireless Communications Technical Committee.
    He is listed as a 2017 Highly Cited Researcher by the Web of Science and a Distinguished Lecturer of the IEEE Communications Society (ComSoc).
    Robert is a Fellow of the Canadian Academy of Engineering and a Fellow of the Engineering Institute of Canada.
    From 2012 to 2015, he served as Editor-in-Chief of the \textsc{IEEE Transactions on Communications}.
    Currently, he is the Chair of the Steering Committee of the \textsc{IEEE Transactions on Molecular, Biological and Multi-Scale Communications}, a Member of the Editorial Board of the Proceedings of the IEEE, a Member at Large of the Board of Governors of ComSoc, and the ComSoc Director of Journals.
\end{IEEEbiography}

\end{document}